\documentclass[a4paper,12pt]{article}
\usepackage{amsfonts, amsmath, amssymb, amsthm}
\usepackage[toc,page]{appendix}
\usepackage[blocks]{authblk}
\usepackage{array}
\usepackage{hhline}
\usepackage{booktabs}
\usepackage[english]{babel}
\usepackage{bm}
\usepackage{dsfont}
\usepackage{eucal}
\usepackage{enumerate}
\usepackage{epsfig}
\usepackage{float}
\usepackage{fdsymbol}
\usepackage{fontenc}
\usepackage{graphicx}
\usepackage[hmargin=2cm,vmargin=3cm]{geometry}
\usepackage[colorlinks,allcolors = blue]{hyperref}
\usepackage{mathrsfs}
\usepackage{mathtools}
\usepackage{multicol}
\usepackage{multirow,bigdelim}
\usepackage{natbib}
\usepackage{rotating}
\usepackage{subfigure}
\usepackage{verbatim}
\usepackage[svgnames]{xcolor}

\usepackage{mdframed}
\usepackage{graphicx}
 \usepackage[normalem]{ulem}

 \definecolor{dodgerblue}{rgb}{0.12, 0.56, 1.0}
\usepackage{color}

\newtheorem{proposition}{Proposition}

\newcommand{\T}{\ensuremath{\mathrm{\scriptscriptstyle T}}}
\usepackage{bm}

\usepackage{amsmath}

\DeclareMathOperator{\E}{E}

\DeclareMathOperator{\cov}{cov}

\usepackage{mathrsfs}

 \let\oldthebibliography=\thebibliography
 \let\oldendthebibliography=\endthebibliography
 \renewenvironment{thebibliography}[1]{%
   \footnotesize
   \oldthebibliography{#1}%
   \setlength{\parskip}{0mm}%
   \setlength{\itemsep}{0mm}%
 }{\oldendthebibliography}

\def\S{\text{S}}

\title{Modeling Interval Trendlines: 
  Symbolic Singular Spectrum Analysis for Interval Time Series
}  
\author[$\dag$]{Miguel~\textsc{de Carvalho}}
\affil[$\dag$]{School of Mathematics, University of Edinburgh, EH9 3FD, UK 
(\href{mailto:miguel.decarvalho@ed.ac.uk}{miguel.decarvalho@ed.ac.uk})}
\author[$\ddag$]{Gabriel~\textsc{Martos}}
\affil[$\ddag$]{Department of Mathematics and Statistics, Universidad Torcuato Di Tella, Argentina
(\href{mailto: gmartos@utdt.edu}{gmartos@utdt.edu})
}

\date{}
\begin{document}
\maketitle 
\vspace{-1cm}
\begin{abstract}\footnotesize
In this article we propose an extension of singular spectrum analysis for interval-valued time series. The proposed methods can be used to decompose and forecast the dynamics governing a set-valued stochastic process. The resulting components on which the interval time series is decomposed can be understood as interval trendlines, cycles, or noise. Forecasting can be conducted through a linear recurrent method, and we devised generalizations of the decomposition method for the multivariate setting. The performance of the proposed methods is showcased in a simulation study. We apply the proposed methods so to track the dynamics governing the Argentina Stock Market (MERVAL) in real time, in a case study that covers the most recent period of turbulence that led to discussions of the government of Argentina with the International Monetary Fund. \\ \ \\ 
\textsc{key words:} Decomposition of interval-valued time series; Interval data; Interval-valued signal; Set-valued stochastic process; Symbolic data analysis; Singular spectrum analysis.
\end{abstract}

\section{Introduction}\label{introduction}
Modeling and forecasting time series with singular spectrum analysis (SSA) has received considerable attention in recent forecasting literature \citep{hassani2013, khan2017, mahmoudvand2018, decarvalho2020}. The rising popularity of the methods stems from the fact that SSA---along with its multivariate version---is naturally tailored for both forecasting and decomposing univariate or multivariate nonstationary time series into a set of principal components, which can be interpreted as trends, cyclical components, or noise. Applications of SSA in practice include predicting inflation dynamics, tracking business cycles, and forecasting industrial production, among other which can be found in the papers above and references therein. For a time series $\mathbf{y} = (a_1, \dots, a_n)$, a key step on which SSA relies is on the the singular value decomposition of a matrix $\bm{Y}$ containing rolling windows of length $l$, that is 
\begin{equation}
  \bm{Y} = 
  \left[
    \begin{array}{cccc}
      a_{1} & a_2 & \cdots & a_k \\ 
      \,a_2 & a_3& \cdots & a_{k+1} \\
      \vdots & \vdots & \ddots & \vdots \\
      \, a_l & a_{l+1}& \cdots & a_{n} \\
    \end{array}
  \right] =  \sum_{i=1}^d \sqrt{\lambda_i} \mathbf{u}_i \mathbf{v}_i^{\T}. 
\label{svd0}
\end{equation}
Here $\lambda_{1} \geq \cdots \geq \lambda_{l}$ and {$\mathbf{u}_1,\ldots,\mathbf{u}_l$} are respectively the eigenvalues and the eigenvectors of $\bm{Y Y^{\T}}$, and $\mathbf{v}_i = \bm{Y}^{\T}{\mathbf{u}_i} /\sqrt{\lambda_i}$ with $d = \max\{i \in \{1,\ldots, l\}: \lambda_i > 0\}$. 

One of the main goals of this paper is to develop SSA methods to model and forecast interval time series, $\mathbf{y} = ([a_1, b_1], \dots, [a_n, b_n])$. There has been an increasing interest on interval time series as can be seen from \cite{gonzalez2012}, \cite{gonzalez2013}, \cite{rodrigues2015}, \cite{lin2016}, and \cite{wang2016}. Interval time series are natural for settings where the interest is on modeling the dynamics of a range of values, such as for instance in financial time series where one is interested in modeling the interval of prices during a trading session (low, high). It is by now well known that naive `midpoint' analyses discard the so-called internal or within variation \citep{le2012}, which in the case of the financial example mentioned earlier corresponds to ignoring intra-day variation. 

A main methodological goal of this article is on developing {univariate and} multivariate singular spectrum analysis methods for interval data that can be used for modeling and forecasting interval time series. The proposed approach will disentangle the dynamics of an interval time series into a sequence of set-valued stochastic processes \citep{kisielewicz2013} that can be interpreted as components underlying trends, regular movements, and noise. The methods proposed in this paper can be readily implemented using the \texttt{R} \citep{rdevelopmentcoreteam2016} package \texttt{ASSA} \citep{dm2018}.

The article is organized as follows. In the next section we introduce symbolic singular spectrum analysis. A simulation study is reported in Section~\ref{simulation}. An application to stock market data can be found in Section~\ref{application}. 

\section{Symbolic Singular Spectrum Analysis}\label{method}
\subsection{Preparations}

Below, the unit of analysis will be an interval-valued time series, $\textbf{y} = ([a_1, b_1], \dots, [a_n, b_n])$. For modeling $\mathbf{y}$ using singular spectrum analysis, we resort to a special type of block matrix to which we refer to as a matrix of ordered pairs.  Below, a matrix $\mathbf{Y}$ is said to be an $k \times l$ matrix of ordered pairs if its elements are ordered pairs, that is
\begin{equation*}
    \mathbf{Y} = \left[ 
      \begin{array}{ccc}
        \ (a_{1,1}, b_{1, 1}) & \ldots & (a_{1, k}, b_{1, k}) \\
        \vdots & \ldots & \vdots \\
        \ (a_{l,1}, b_{l, 1}) & \ldots & (a_{l, k}, b_{l, k}) \\
      \end{array}
    \right].
\end{equation*}
%\begin{equation*}
%    \mathbf{Y} = \left[ 
%      \begin{array}{ccc}
%        \ [a_{1,1}, b_{1, 1}] & \ldots & [a_{1, k}, b_{1, k}] \\
%        \vdots & \ldots & \vdots \\
%        \ [a_{l,1}, b_{l, 1}] & \ldots & [a_{l, k}, b_{l, k}] \\
%      \end{array}
%    \right].
%\end{equation*}
Throughout, sums and differences between such matrices should be understood as their respective pointwise Minkowski-type counterparts, respectively defined as 
 \begin{equation*}
 \resizebox{\textwidth}{!}{$ \textbf{A} + \textbf{B} = \left[
    \begin{array}{ccc}
      \ (a_{1, 1} + c_{1, 1}, b_{1, 1} + d_{1, 1}) & \cdots & (a_{1, k} + c_{1, k}, b_{1, k} + d_{1, k}) \\ 
      \vdots & \ddots & \vdots \\
      \ (a_{l, 1} + c_{l, 1}, b_{l, 1} + d_{l, 1}) & \cdots & (a_{l, k} + c_{l, k}, b_{l, k} + d_{l, k})
    \end{array}
     \right], \quad
 \mathbf{A} - \mathbf{B} = \left[
    \begin{array}{ccc}
      \ (a_{1, 1} - c_{1, 1}, b_{1, 1} - d_{1, 1}) & \cdots & (a_{1, k} - c_{1, k}, b_{1, k} - d_{1, k}) \\ 
      \vdots & \ddots & \vdots \\
      \ (a_{l, 1} - c_{l, 1}, b_{l, 1} - d_{l, 1}) & \cdots & (a_{l, k} - c_{l, k}, b_{l, k} - d_{l, k})
    \end{array}
     \right], $
 }
 \end{equation*}
where
\begin{equation*}
\mathbf{A} = \left[
\begin{array}{ccc}
  \ (a_{1, 1}, b_{1, 1}) & \cdots & (a_{1, k}, b_{1, k}) \\ 
  \vdots & \ddots & \vdots \\
  \ (a_{l, 1}, b_{l, 1}) & \cdots & (a_{l, k}, b_{l, k})
\end{array}
\right], \qquad 
\mathbf{B} = 
\left[
\begin{array}{ccc}
  \ (c_{1, 1}, d_{1, 1}) & \cdots & (c_{1, k}, d_{1, k}) \\ 
  \vdots & \ddots & \vdots \\
  \ (c_{l, 1}, d_{l, 1}) & \cdots & (c_{l, k}, d_{l, k}) \\
\end{array}
\right]. 
\end{equation*}
{The components of the resulting matrices $\textbf{A} + \textbf{B}$ and $\textbf{A} - \textbf{B}$ 
can be naturally mapped into interval data via the set-valued function  
$\phi(x,y)=[\min\{x,y\},\max\{x,y\}]$; note that $\phi(a_{i,j}+ c_{i,j},b_{i,j}+d_{i,j})= [a_{i,j}+c_{i,j},b_{i,j}+d_{i,j}]$ and $\phi(a_{i,j}- c_{i,j},b_{i,j}-d_{i,j}) = [\min\{a_{i,j}- c_{i,j},b_{i,j}- d_{i,j}\},\max\{a_{i,j} - c_{i,j},b_{i,j}- d_{i,j}\}]$ for $1\leq i \leq l,$ and $1\leq j\leq k$ respectively.
} To compute the norm of $\mathbf{Y}$, we use the following variant of the Frobenious norm:
\begin{equation}\label{cnorm}
  \|\mathbf{Y}\|_{\text{C}} = \frac{1}{\sqrt{2}} \left\{\sum_{i = 1}^l \sum_{j = 1}^k (a_{i, j}^2 + b_{i, j}^2)\right\}^{1 / 2},
\end{equation}
for a matrix of ordered pairs $\mathbf{Y} = \{(a_{i, j}, b_{i, j})\}$; note that if $a_{i, j} = b_{i, j}$ we recover the standard Frobenious norm. Of particular interest for our developments is the class of what we will refer to as Hankel matrices of ordered pairs. An $l \times k$ matrix of ordered pairs is said to be an Hankel matrix (of ordered pairs) if its elements coincide on the antidiagonals $i + j = s$ for any $2 \leq s \leq l + k$. The notation $\mathscr{H}_{l, k}$ will be used throughout to denote the space of all $l \times k$ Hankel matrices of ordered pairs. 

\subsection{Interval-Valued Singular Spectrum Analysis (IVSSA)}\label{sec:ivssa}
Let $\mathbf{y} = ([a_1, b_1], \dots, [a_n, b_n])$ be an
interval-valued time series.  Interval-valued singular spectrum
analysis (IVSSA), to be proposed below, can be regarded as an
extension of singular spectrum analysis \citep{golyandina2013} to be used for decomposing an
interval-valued time series into components, and for learning about interval trendlines from data.
IVSSA entails two phases, namely decomposition and reconstruction, and each of these phases includes two steps. The decomposition includes the steps of embedding and symbolic singular value decomposition, which we discuss below.  \\

\noindent \textsc{Embedding.}~{IVSSA starts by organizing the original interval-valued time series of interest, $\mathbf{y} = ([a_1, b_1], \dots, [a_n, b_n])$, into a trajectory matrix $\bm{Y}$, i.e., a matrix of ordered pairs whose columns consist of rolling windows of length $l$, as follows
\begin{equation}
  \bm{Y} = 
  \left[
    \begin{array}{cccc}
      (a_{1}, b_1) & (a_2, b_2) & \cdots & (a_k, b_k) \\ 
      \,(a_2, b_2) & (a_3, b_3)& \cdots & (a_{k + 1}, b_{k + 1}) \\
      \vdots & \vdots & \ddots & \vdots \\
      \, (a_l, b_l) & (a_{l + 1}, b_{l + 1})& \cdots & (a_{n}, b_{n}) \\
    \end{array}
  \right], 
  \label{traj_matrix}
\end{equation}
\noindent where $l$ is set by the user and $k=n-l+1$. {Here $l$ is a signal--noise separation parameter that plays a similar role to that of the bandwidth in nonparametric regression. Since
all elements over the diagonal $i + j = \text{const}$ are equal, then $\bm{Y} \in \mathscr{H}_{l, k}$.}\\

\noindent \textsc{Symbolic singular value decomposition.}~{In the second step we perform a symbolic singular value decomposition of the trajectory matrix resorting to the so-called covariance matrix for symbolic data, as defined in \cite{billard2007} and \cite{le2012}. Let $\lambda_{1}^\S\geq \cdots \geq \lambda_{l}^\S$ be the eigenvalues and {$\mathbf{u}_1^{\S},\ldots,\mathbf{u}_l^{\S}$} be the  eigenvectors corresponding to matrix $\bm{S}\in \mathbb{R}^{l\times l}$, with entries given by 
\begin{equation*}
 [\bm{S}]_{jj'} = s_{jj'} = \frac{1}{6} \sum_{i=1}^k \big\{2a_{j+i-1}a_{j'+i-1} +a_{j+i-1}b_{j'+i-1} + b_{j+i-1}a_{j'+i-1} +2b_{j+i-1}b_{j'+i-1}\big\}, 
\end{equation*}
where $s_{jj'}$, for $1\leq j \leq l$ and $1\leq j' \leq l$, up to a constant, represent the estimated covariance between interval data in rows $j$ and $j'$ on $\bm{Y}$ \citep[][Eq.~3]{billard2012}. %Notice that $s_{jj}$ is, up to a constant, the estimated variance corresponding to the interval data in the $j$--th row  of $\bm{Y}$--\citep[][p.~416, Eq.(2.3)]{le2012}--. 
We resort on the eigenvectors and eigenvalues of $\bm{S}$ to decompose the trajectory matrix $\bm{Y}$ as follows
\begin{equation}
  \bm{Y} = \sum_{i=1}^d \bm{Y}_i, %\sqrt{\lambda_i^{\S}} {\mathbf{w}_i^{\S} (\mathbf{v}_i^{\S})^{\T}},
\label{svd1}
\end{equation}
where $\bm{Y}_i = \sqrt{\lambda_i^{\S}}\mathbf{u}_i^{\S}(\mathbf{v}_i^{\S})^{\T}$, $\mathbf{v}_i^{\S} = \bm{Y}^{\T}{\mathbf{u}_i^{\S}} /\sqrt{\lambda_i^{\S}}$ and $d = \max\{i \in \{1,\ldots, l\}: \lambda_i^{\S} > 0\}$. {Notice that 
$$ \mathbf{v}_i^{\S} = \frac{1}{\sqrt{\lambda_i^{\S}}} \left[\left(\sum_{j=1}^lu_{i,j}^{\S}a_{j},\sum_{j=1}^lu_{i,j}^{\S}b_{j}\right),\dots, \left(\sum_{j=1}^lu_{i,j}^{\S}a_{j+k-1},\sum_{j=1}^lu_{i,j}^{\S}b_{j+k-1}\right)\right]^{\T};$$
thus, the resulting matrices $\bm{Y}_i$ 
are matrices of ordered pairs, for $i=1,\dots,d$.} % corresponding to interval data. 
Next we discuss the reconstruction phase, which involves the steps of grouping components and diagonal averaging.\\

\noindent \textsc{Grouping.}~Not all 
terms in Equation~\eqref{svd1} contain relevant
information on the interval trendline, and hence we retain only a subset $I \subset \{1, \ldots, d\}$ to compute  $\bm{Y}_I = \sum_{i \in I} \bm{Y}_i$. The goal of this step is on disentangling the signal from noise, assuming that   $\bm{Y} = \bm{Y}_I + \boldsymbol{\varepsilon}$, being $\bm{Y}_I$ the signal and $\boldsymbol{\varepsilon} = \sum_{i \not \in I} \bm{Y}_i$ the noise on the data. To learn about $I$, in Appendix~A.2 we show how the periodogram-based method of  \cite{decarvalho2020} can be extended to an interval-valued time series context by devising a periodogram for interval-valued time series. The proposed extension is based on the analysis of the periodogram of an interval-valued time series of residuals (termed below as Hausdorff residuals). The strengths and limitations with such periodogram-based approach will be numerically examined in Section~\ref{simulation}.\\ % Regardless of the method chosen to compute $I$, we recommend carrying out a sensitivity analysis on this parameter so to assess the robustness of the model}.\\

\noindent \textsc{Diagonal averaging.}~
In this step we average over all the elements of the (anti)diagonal $i + j = \textit{const}$ of $\bm{Y}_I$ so to obtain an Hankel matrix of ordered pairs, from where our interval trendline indicator results. The following proposition provides the formal justification for this step; see Appendix~A.1 for a proof.
\begin{proposition}\label{daver}
Let $\mathscr{H}_{l, k}$ be the space of all $l \times k$ Hankel matrices of ordered pairs. Let $\mathbf{Y} = \{(a_{i, j}, b_{i, j})\}$ be an $l \times k$ matrix of ordered pairs. Then,
\begin{equation*}
  \mathbf{H}^* = \{(\alpha_{i, j}^*, \beta_{i, j}^*)\}= 
  \arg \min_{\mathbf{H} \in \mathscr{H}_{l, k}} 
  \|\mathbf{Y} - \mathbf{H} \|_{\emph{C}}, 
\end{equation*}
where $(\alpha_{i, j}^*, \beta_{i, j}^*) = (1 / n_s \sum_{i + j = s} a_{i, j}, 1 / n_s \sum_{i + j = s} b_{i, j})$, with $n_s$ denoting the number of $(i, j)$ such that $i + j = s$, with $i \in \{1, \dots, l\}$ and $j \in \{1, \dots, k\}$.
\end{proposition}
\noindent Thus, following Proposition~\ref{daver}, we construct our interval trendline indicator by averaging the matrix of ordered pairs $\bm{Y}_I$ over the antidiagonals $i + j = s$. Let $(a^I_{ij},b^I_{ij}) = [\mathbf{Y}_I]_{ij}$ then for $s = 2$, ${\widetilde{y}_1}=\phi(a_{11}^{I}, b_{11}^{I})$; $s = 3$, yields ${\widetilde{y}_2}=\phi(a_{1 2}^{I} + a_{21}^{I},b_{1 2}^{I} + b_{22}^{I})/ 2$; $s = 4$, yields ${\widetilde{y}_3}=\phi(a_{13}^{I} + a_{2 2}^{I} + a_{31}^{I},b_{13}^{I} + b_{22}^{I} + b_{31}^{I})/ 3$; \textit{etc}. Extending this simple construct, we build our interval trendline indicator through the map 
\begin{equation}
\widetilde{\mathbf{y}} = \{[\widetilde a_t, \widetilde b_t]\}_{t=1}^n = \overline{\mathbb{D}}(\bm{Y}_I) \equiv 
\left\{\frac{1}{n_2}\phi\left(\sum_{i + j = 2}a_{i j}^{I}, \sum_{i + j = 2}b_{i j}^{I}\right), \ldots, 
\frac{1}{n_{n + 1}}\phi\left(
\sum_{i + j = n + 1} 
a_{ij}^{I}, \sum_{i + j = n + 1}b_{ij}^{I}\right) \right\}, 
\label{diag.aver}
\end{equation}
with $n_s$ denoting the number of $(i, j)$ such that $i + j = s$, with $i \in \{1, \dots, l\}$ and $j \in \{1, \dots, k\}$. 
\subsection{Selected Comments on Forecasting with IVSSA}\label{forecasting}
\label{forecasting}
Similarly to SSA for time series analysis, forecasting can be here conducted via a recurrent forecasting algorithm \citep[Chapter~3]{golyandina2013}. The recurrent forecasting method relies on an autoregressive-type 
assumption that specifies that the i$th$ interval observation $y_i=[a_i,b_i]$ is a combination of the preceding $l-1$ observations, so that for all $i \geq l$ it holds that
\begin{equation}\label{recurrent}
  [a_{i},b_{i}]  =\phi( \alpha_1(a_{i-1},b_{i-1}) + \dots+ \alpha_{l-1}(a_{i-l+1},b_{i-l+1})) = \phi\left(\sum_{j=1}^{l-1}\alpha_j a_{i-j},\sum_{j=1}^{l-1}\alpha_jb_{i-j} \right),
\end{equation}
where $\boldsymbol\alpha = (\alpha_1,\dots,\alpha_{l-1})$ is a vector of coefficients. The specification in \eqref{recurrent} can then be used for forecasting. For example, the one-step forecast, $[\widehat{a}_{n+1},\widehat{b}_{n+1}]$, is a combination of the most recent $l-1$ interval-valued signals, that is 
$$[\widehat{a}_{n+1},\widehat{b}_{n+1}]  = \phi( \alpha_1(\widetilde{a}_{n},\widetilde{b}_{n}) + \dots+ \alpha_{l-1}(\widetilde{a}_{n-l+2},\widetilde{b}_{n-l+2})) = \phi\left(\sum_{j=1}^{l-1}\alpha_j \widetilde{a}_{n+1-j},\sum_{j=1}^{l-1}\alpha_jb_{n+1-j} \right).$$ 
And the out-of-sample forecasts corresponding to the time periods $n+2$, $n+3$, $\dots$, are obtained using the previous formula recursively. 
The question of forecasting via \eqref{recurrent} boils down to obtaining the vector $\boldsymbol \alpha$, which can be retrieved from the symbolic singular value decomposition via \citet[][Proposition~1]{golyandina2001}. Following \cite{rodrigues2013} the vector $\boldsymbol \alpha$ can be computed as follows 
\begin{equation}\label{forecast}
  \boldsymbol \alpha = \frac{\bm{M}[\bm{\Pi}_1\odot(\bm{\pi}_1\otimes \bm{1}_{l-1})] \bm{1}_{m}}{1-\|\bm{\pi}_{1}\|^2},
\end{equation}
where $\| \cdot \|$ is the Euclidean norm, $\odot $ and $\otimes $ are the Hadamard and tensor Kronecker products, respectively,   and $\bm{M} \in \mathbb{R}^{(l-1)\times(l-1)}$ is an antidiagonal matrix with ones in the main antidiagonal. In addition, $\bm{\Pi}_1 \in \mathbb{R}^{(l-1)\times m}$ is a matrix composed by the first $(l-1)$ components of the $m$ eigenvectors associated to  signal, whereas $\bm{\pi}_1 \in \mathbb{R}^{1\times m}$ contains the last components of those eigenvectors. 

The next section will consider multivariate extensions of IVSSA. 

\subsection{Multivariate Extensions}\label{sec:mvssa}
Suppose now that we observe $D$ interval time series, $\{\mathbf{y}_i = [a_{ij},b_{ij}]\}_{j=1}^{N_i}$, where $N_i$ denotes the series $i$th length, for $i=1,\dots,D$. Multivariate IVSSA (MIVSSA) entails a similar course of action as that described in Section~\ref{sec:ivssa}. To streamline the discussion we assume that $N_1=\dots=N_D\equiv N$ and $l_1=\dots =l_D \equiv l$, but all steps below can be easily adapted otherwise. \medskip

\noindent \textsc{Embedding and Singular value decomposition.}~{ Let $\bm{Y}_i \in \mathcal{H}_{l,k}$ be the interval trajectory matrix %--see Equation~\eqref{traj_matrix}-- 
corresponding to the $i$th series; we consider either of the two stacked trajectory matrix:
\begin{equation}
    \bm{Y}_V =  \left[ \begin{array}{c}
\bm{Y}_1  \\ \vdots \\ \bm{Y}_D
  \end{array} \right]
  , \quad  \text{ or } \quad  \bm{Y}_H =  \left[ \begin{array}{c}
 \bm{Y}_1~~~\cdots ~~~\bm{Y}_D
   \end{array} \right],
\label{traj_matrix2}
\end{equation}
where $\bm{Y}_V\in \mathcal{H}_{lD,k}$ stand for vertical stack and  $\bm{Y}_H\in \mathcal{H}_{l,kD}$ for horizontal stack.} %vis-a-vis 
In regard to the stacking strategy, we consider the eigen--pairs corresponding to matrices: 
\begin{equation} 
\bm{S}_V=
  \left[\begin{array}{ccc}
    \bm{S}_{11}   & \cdots & \bm{S}_{1 D}\\
    \vdots  & \ddots & \vdots \\
         {\bm{S}}_{D1}  & \cdots & \bm{S}_{DD} 
                   \end{array}\right], \qquad \bm{S}_H= \sum_{i=1}^D \bm{S}_{ii},
\end{equation}  
where $[\bm{S}_{ii'}]_{jj'}= \sum_{q=1}^k \{2a_{i,j+q-1}a_{i',j'+q-1} +a_{i,j+q-1}b_{i',j'+q-1} + b_{i,j+q-1}a_{i',j'+q-1} +2b_{i,j+q-1}b_{i',j'+q-1}\}/6,$
%$$[\bm{S}_{ii'}]_{jj'}= \frac{1}{6} \sum_{q=1}^k [2a_{i,j+q-1}a_{i',j'+q-1} +a_{i,j+q-1}b_{i',j'+q-1} + b_{i,j+q-1}a_{i',j'+q-1} +2b_{i,j+q-1}b_{i',j'+q-1}],$$
for $j,j'=1,\dots,l$ and $i,i'=1,\dots,D$
%$1\leq j \leq l$, $1\leq j' \leq l$, $1\leq i \leq D$ , $1\leq i' \leq D$. 
. For vertical staking, the elements in the diagonal of $\bm{S}_V$ correspond to different interval covariance matrices obtained when applying IVSSA on each interval time series separately. Considering %that the 
vertical stacking and denoting as $\{(\lambda_{1}^{_\text{V}},\mathbf{u}_{1}^{_\text{V}}), \ldots,(\lambda_{lD}^{_\text{V}},\mathbf{u}_{lD}^{_\text{V}})\}$ the eigen--pairs of $\bm{S}_V$, then ${\bm{Y}}_V = \sum_{i=1}^{d} \bm{Y}_i$, where $\bm{Y}_i = \sqrt{\lambda_i^{_\text{V}}} \mathbf{u}_i^{_\text{V}} (\mathbf{v}^{_\text{V}}_i)^T$ and $\mathbf{v}_i^{_\text{V}} = \bm{Y}_V^T\mathbf{u}_i^{_\text{V}} /\sqrt{\lambda_i^{_\text{V}}}$, for $i = 1, \dots, d$ where $d = \max\{i \in \{1,\ldots, lD\}: \lambda_i^{_\text{V}} > 0\}$.\\

 \noindent \textsc{Grouping and Diagonal averaging.}~Not all terms in the decomposition of $\bm{Y}_V$ contain information about the signal, hence we retain a subset $I\subset \{1, \ldots, d\}$ %of the estimated ERC, 
so to compute $\bm{Y}_I=  \sum_{i \in I} \bm{Y}_i$.
Averaging over all the elements of the (anti)diagonal $i + j = \textit{const}$ of $\bm{Y}_I$ yields an interval Hankel matrix, from where our interval trendline indicators results. We allow for each trendline to be constructed from a different number of components, that is 
\begin{equation}\label{mtrendline}
  [\widetilde{\mathbf{y}}_1~\cdots~\widetilde{\mathbf{y}}_D]  = [\mathbb{D}(\bm{Y}_{I_{1}}^1)~\cdots~\mathbb{D}(\bm{Y}_{I_{D}}^D)].
\end{equation}
where 
\begin{equation*}
  \left[
    \begin{array}{c}  
      \bm{Y}_{I_{1}}^1 \\
      \vdots \\
      \bm{Y}_{I_{1}}^D \\
    \end{array}\right]
  = \sum_{i \in I_1} \bm{Y}_i \, ,
  \dots 
  , 
  \left[
    \begin{array}{c}  
      \bm{Y}_{I_{D}}^1 \\
      \vdots \\
      \bm{Y}_{I_{D}}^D \\
    \end{array}\right]
  = \sum_{i \in I_D} \bm{Y}_i.
\end{equation*}
%To learn about the sets $I_{1},\dots,I_{D}$, we use the targeted grouping approach described in Appendix~A.2. 
%and $\mathbb{D}$ defined as in Equation~\eqref{diag.aver}.
Note that \eqref{mtrendline} is thus a multivariate version of \eqref{diag.aver}. Next, we assess the finite sample performance of the proposed methods in a simulation study.

\section{Simulation Study}\label{simulation}

\subsection{Data Generating Scenarios and Preliminary Experiments}\label{oneshot}
In this section we assess the performance of the proposed methods via a simulation study. A Monte Carlo study will be presented in Section~\ref{mc}; for now we concentrate on discussing the data generating processes from which the data are simulated, and on illustrating a fit from the proposed methods on a single run experiment with $n = 100$. We consider the following interval-valued  data generating processes:
\begin{equation}\label{xy}
x_t = [\mu^x_{t} + \varepsilon_t^x, \mu^x_{t} + 2 + \varepsilon_t^x], 
\qquad y_t = [\mu^y_{t} + \varepsilon_t^y, 2(\mu^y_{t}+1) + \varepsilon_t^y],
 \end{equation}
where $t\in T = \{2i\pi/n\}_{{i=1}}^n$, $\mu^x_{t} = 8 + t + \sin(\pi t)$, $\mu^y_{t}  = \sqrt{t} + \cos(\pi t/2)$. Here,  $\varepsilon_t^x$ and $\varepsilon_t^y$ are zero mean normally distributed errors %errors with a Mat\'ern function,
%$$
%\text{Cov}(X_t,X_s) = \text{Cov}(Y_t,Y_s) \equiv 
%c(t,s)= \frac{2^{1-\nu}}{ %\Gamma(\nu)}\Big(\sqrt{\frac{2\nu}%{\theta}}\|t-s\|\Big)^\nu K_\nu\Big(\sqrt{\frac{2\nu}{\theta}}\|t-s%\|\Big);
%$$
%here, $K_\nu$ is a Mat\'ern family kernel and $\nu=1/2$ and $\theta=1/10$ are respectively the scale and smoothness parameters. The cross-covariance 
with covariance function given by
\begin{equation*}
\Sigma(s,t)\equiv \begin{bmatrix}
\text{Cov}(\varepsilon_t^x,\varepsilon_s^x) & \text{Cov}(\varepsilon_t^x,\varepsilon_s^y)\\
\text{Cov}(\varepsilon_t^y,\varepsilon_s^x)& \text{Cov}(\varepsilon_t^y,\varepsilon_s^y)
\end{bmatrix} =
 \delta(t-s)\begin{bmatrix}
\sigma^2 & \rho\\\
\rho & \sigma^2
\end{bmatrix},
\end{equation*}
where $\delta(0)=1$ and $\delta(t-s)=0$ for $t\neq s$. In Scenario~A we set $\rho = 0$ and $\sigma^2=1$, thus $\{x_t\}$ and $\{y_t\}$ are independent interval-valued processes, and in Scenario~B we set $\rho = 1/2$ and $\sigma^2=1$, leading to dependent interval-valued processes.

\begin{figure}[H]
\centering
\textbf{Scenario~A}\\
\includegraphics[scale = 0.67]{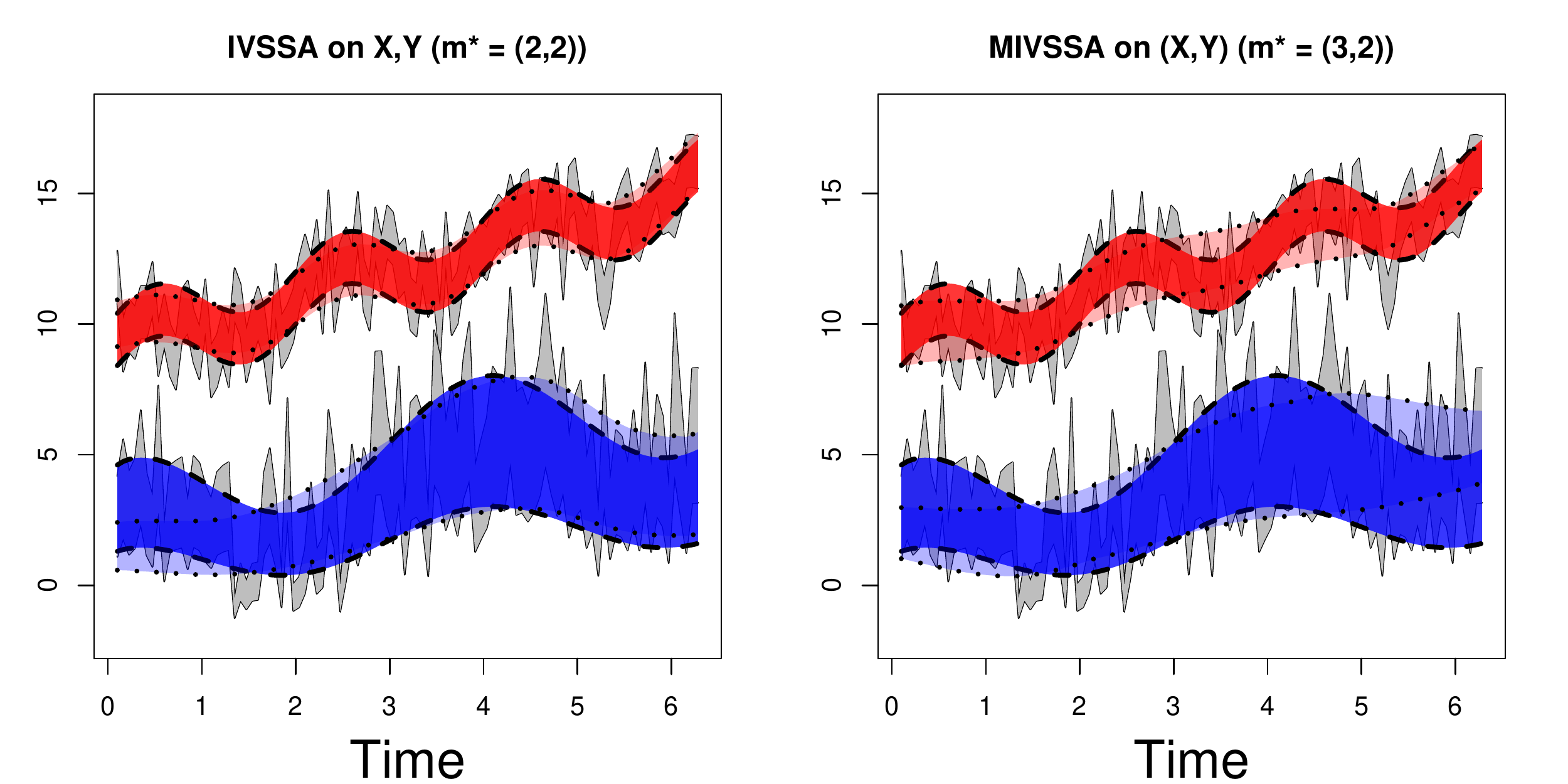}\\
\textbf{Scenario~B}\\
\includegraphics[scale = 0.67]{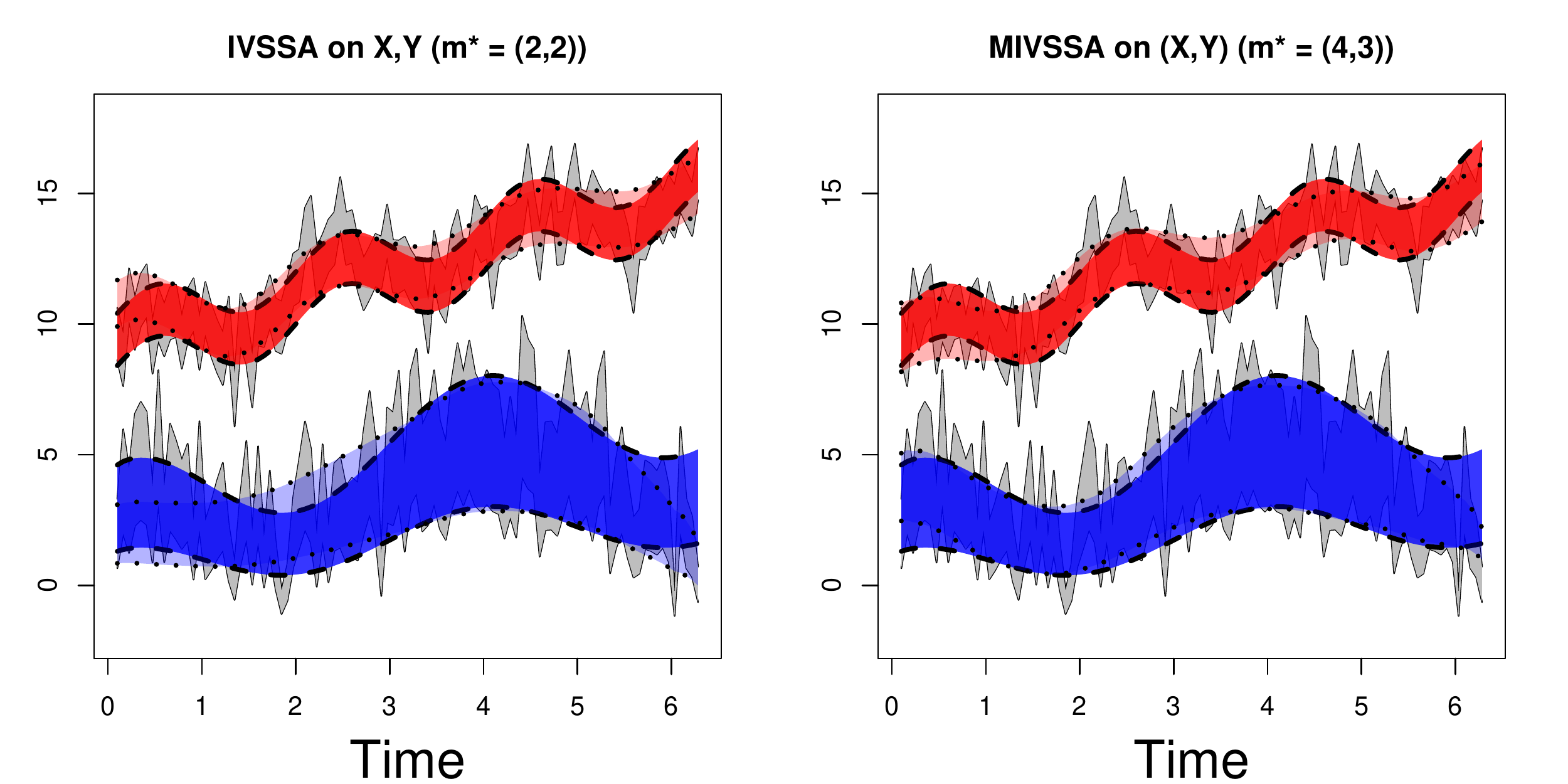}
\caption{\footnotesize One shot experiments for Scenario~A and~B. The solid gray areas corresponds to raw interval data for $\{x_{t}\}$ (left), $\{y_{t}\}$ (middle), and  $\{(x_{t}, y_{t})\}$ (right). Conditional means $[\mu_t^X,\mu_t^X+2]$ and $[\mu_t^Y,2(\mu_t^Y+1)]$ and trendline estimators are depicted in solid and transparent red and blue respectively. \label{ilust}}
\end{figure}

{The processes $\{x_t\}$ and $\{y_t\}$ will be used to illustrate all versions of the proposed method, namely: Interval-Valued Singular Spectrum Analysis (IVSSA; Section~\ref{sec:ivssa}) as well as its multivariate extensions (h-MIVSSA and v-MIVSSA; Section~\ref{sec:mvssa}).} In Figure~\ref{ilust} we present one instance of an interval trendline estimate yield using our methods corresponding to a one shot experiment for Scenarios~A--B. As it can be seen from Figure~\ref{ilust} our methods closely track the true interval means of both processes for Scenarios~A--B; of course such finding should be regarded as tentative, as this is the outcome of a single run experiment, but the same inquiry will be revisited in Section~\ref{mc} through the lenses of a Monte Carlo simulation study.

Some comments on the selection of $(m, l)$ over our numerical experiments are in order. {As anticipated in Section~\ref{method}, to learn about $m$ we adapt the periodogram-based approach in \cite{decarvalho2020} to an interval-valued setting; details on the latter are available from Appendix~A.1.} Keeping in mind theoretical results on the window length achieving maximum rank \cite[][Section~4.1]{hassani2013}, we consider  $l = \lceil (n + 1) / (D + 1)\rceil \label{lrule}$ in the case of vertical stacking and $l = \lceil D(n + 1) / (D + 1)\rceil$ in the case of horizontal stacking, where $\lceil \cdot \rceil$ is the ceiling function.
%
%
%Keeping in mind theoretical considerations in \citet[][Section~4.1]{hassani2013a} we set $l =\lceil (n + 1)/2 \rceil$ for IVSSA,\footnote{\blue{Gabriel: Can you please complete keeping in mind the rules for vertical and horizontal?}} where $\lceil \cdot \rceil$ is the ceiling function. 

 % so to assess the reliability of the 
% latter finding over different simulated data sets.  

\begin{figure}
  \centering
  \hspace{-1.5cm}
   \textbf{Scenario~A}
   \includegraphics[scale = 0.67]{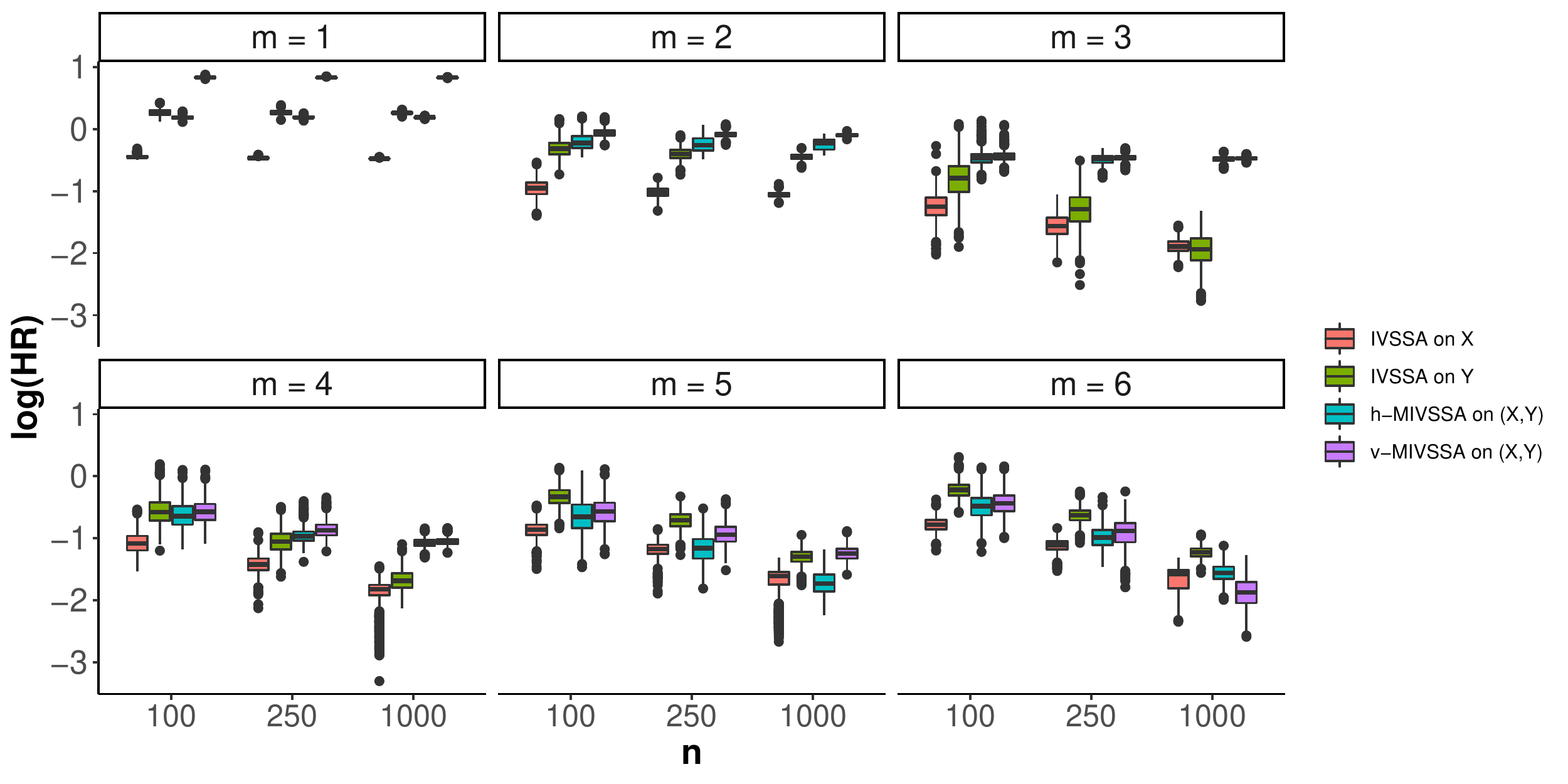} \\ \ \\ 
   \hspace{-1.5cm}
   \textbf{Scenario~B}
   \includegraphics[scale = 0.67]{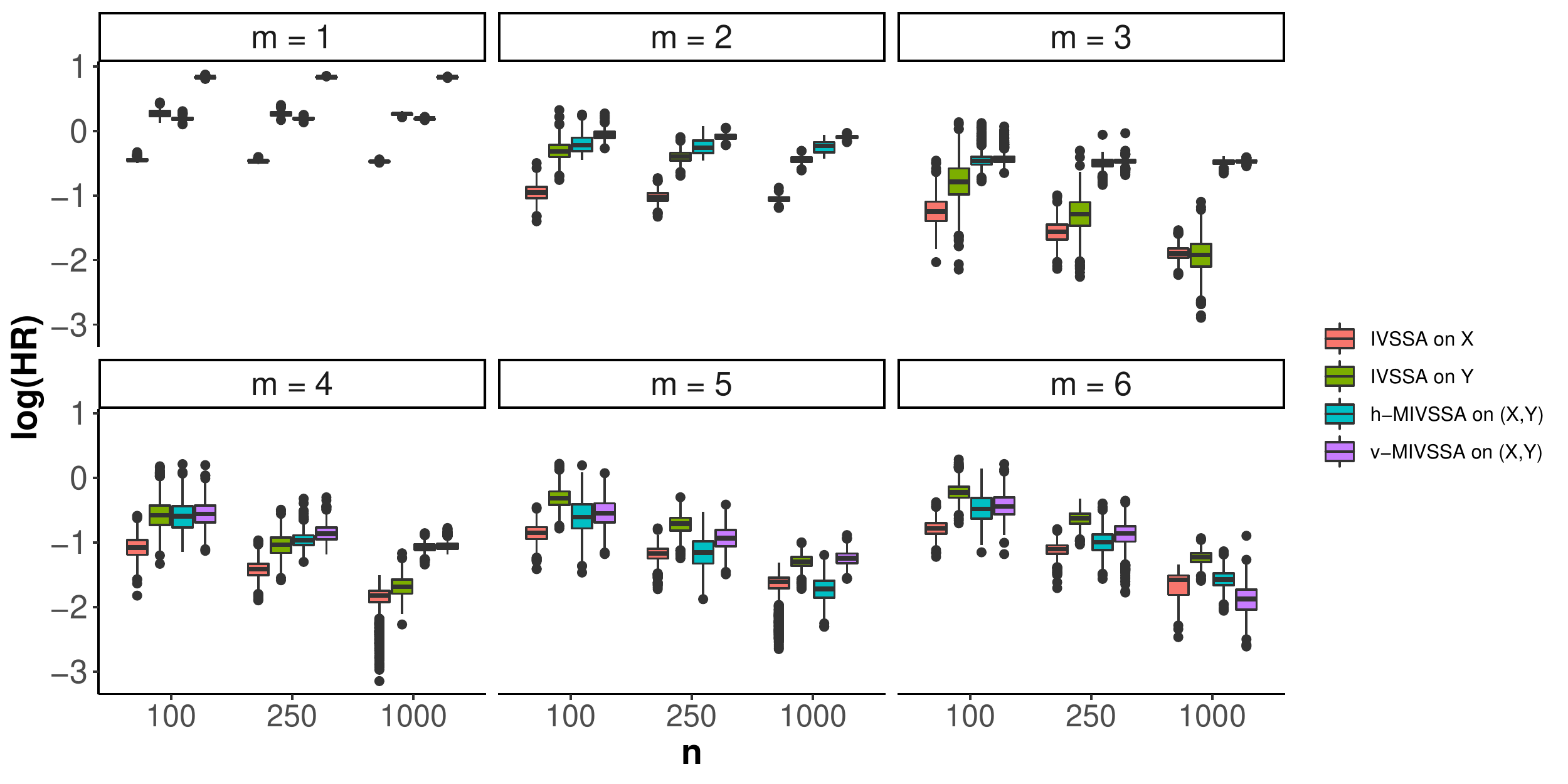}
   % \caption{Scenario A: In sample HE distribution for different values of $m$ and $n$.\label{sceA}}
   \caption{\footnotesize Monte Carlo simulation study Hausdorff residuals 
     (HR). Boxplots of HR for Scenarios~A and B over different values of
     $m$ and $n$.\label{sce}}
\end{figure}

\subsection{Monte Carlo Simulation Study}\label{mc}%\footnote{\textbf{\blue{This section was substantially revised. Please read and let me know next time about your comments.}}}
We now report the main findings of a Monte Carlo simulation experiment based on the data generating processes described in Section~\ref{oneshot}; here, we consider $S=1\,000$ Monte Carlo simulations. For Scenarios~A and B we consider the sample sizes $n = 100, 250, 1\,000$, and allow for the number of ERC to be retained to be $m = 1, \dots, 6$. Since the processes under study are set-valued, we assess performance using the Hausdorff distance between the mean set-valued process ($\E(x_t)$) and the estimated interval trendline ($\widetilde{x}_t$), that is 
$$D_{\mathrm {H} }(\E(x_t) , \widetilde{x}_{t})= \max \{|\E(a_{t})-\widetilde{a}_{t}|,|\E(b_{t})-\widetilde{b}_{t}| \},$$ with $\{x_t \equiv [a_t,b_t]\}_{t}$ and $\{\widetilde x_t \equiv [\widetilde a_t, \widetilde b_t]\}_{t}$.
More specifically, we compute the average Hausdorff residuals (HR) here defined as $$\text{HR} = \frac{1}{n} \sum_{t}D_{\mathrm {H} }(\E(x_t) ,\widetilde{x}_{t}).$$
Figure~\ref{sce} depicts side-by-side boxplots of HEs for Scenarios A and B; in the Supplementary Material, we also report the Monte Carlo mean HE for all the sample sizes and ERCs in this study.

\begin{figure}[H]
  \centering
%  \textbf{Scenario~A} \\ 
  \includegraphics[scale=0.33]{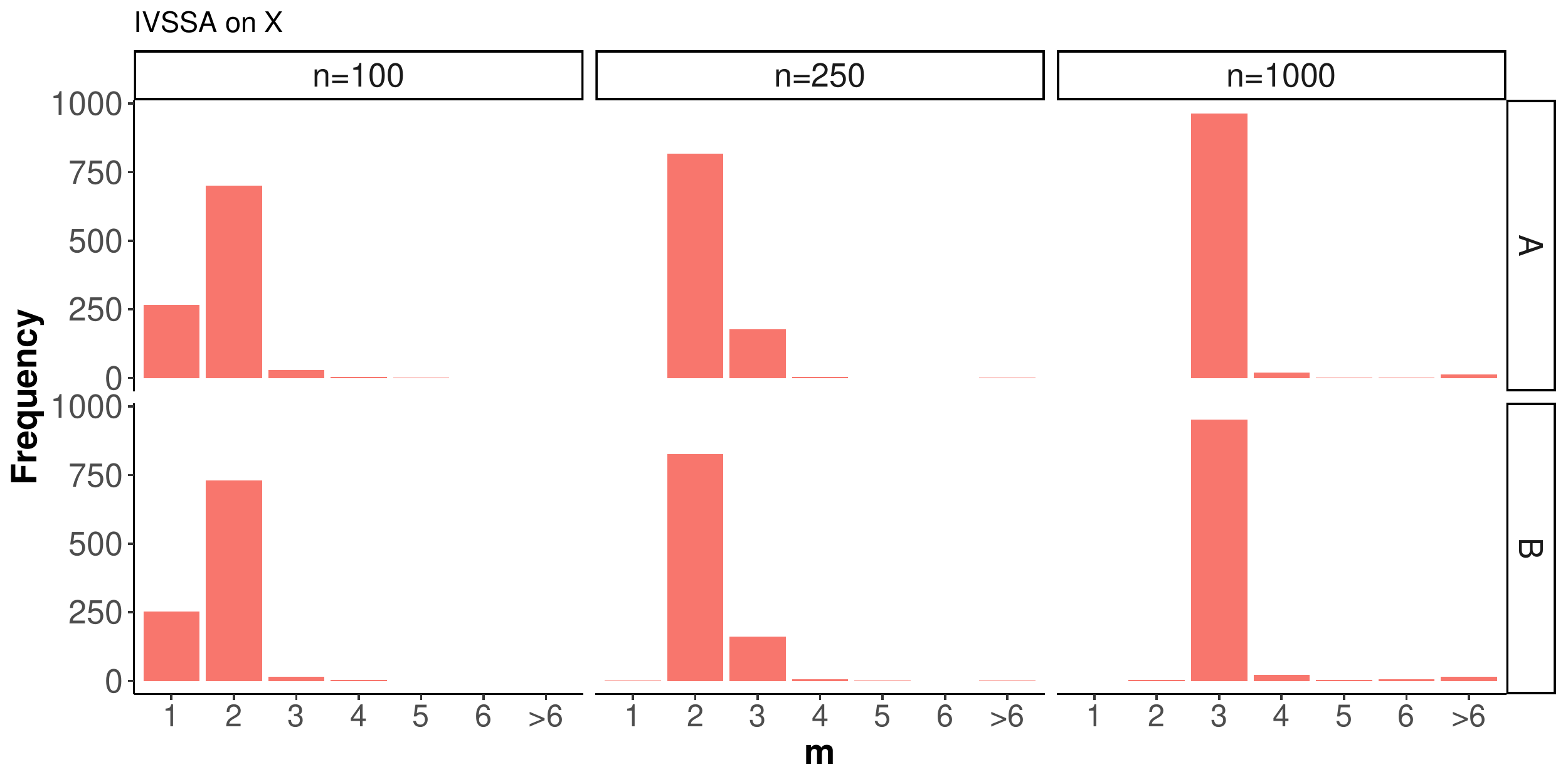}
  \includegraphics[scale=0.33]{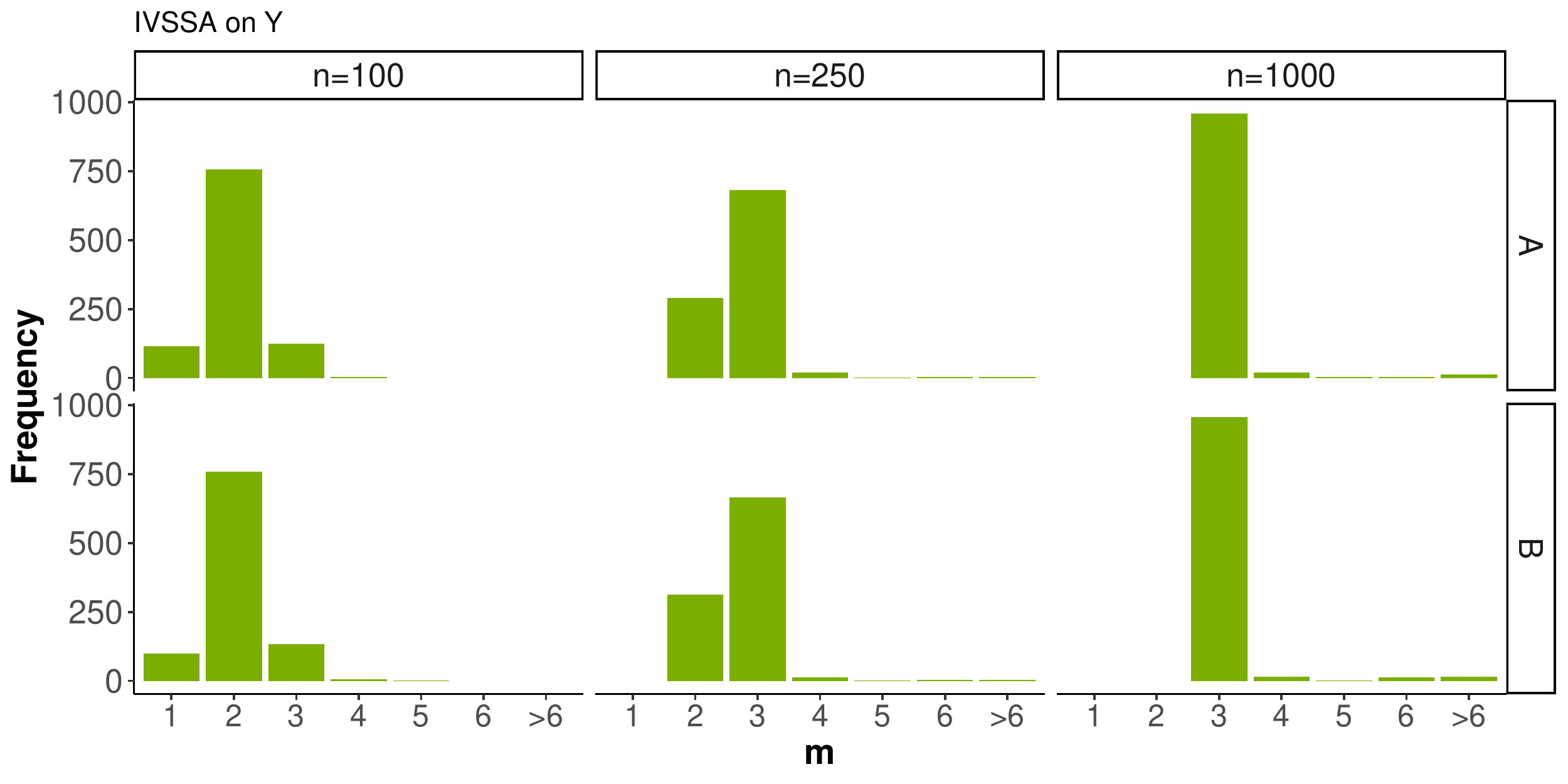} \\
 %   \textbf{Scenario~B} \\
\includegraphics[scale=0.33]{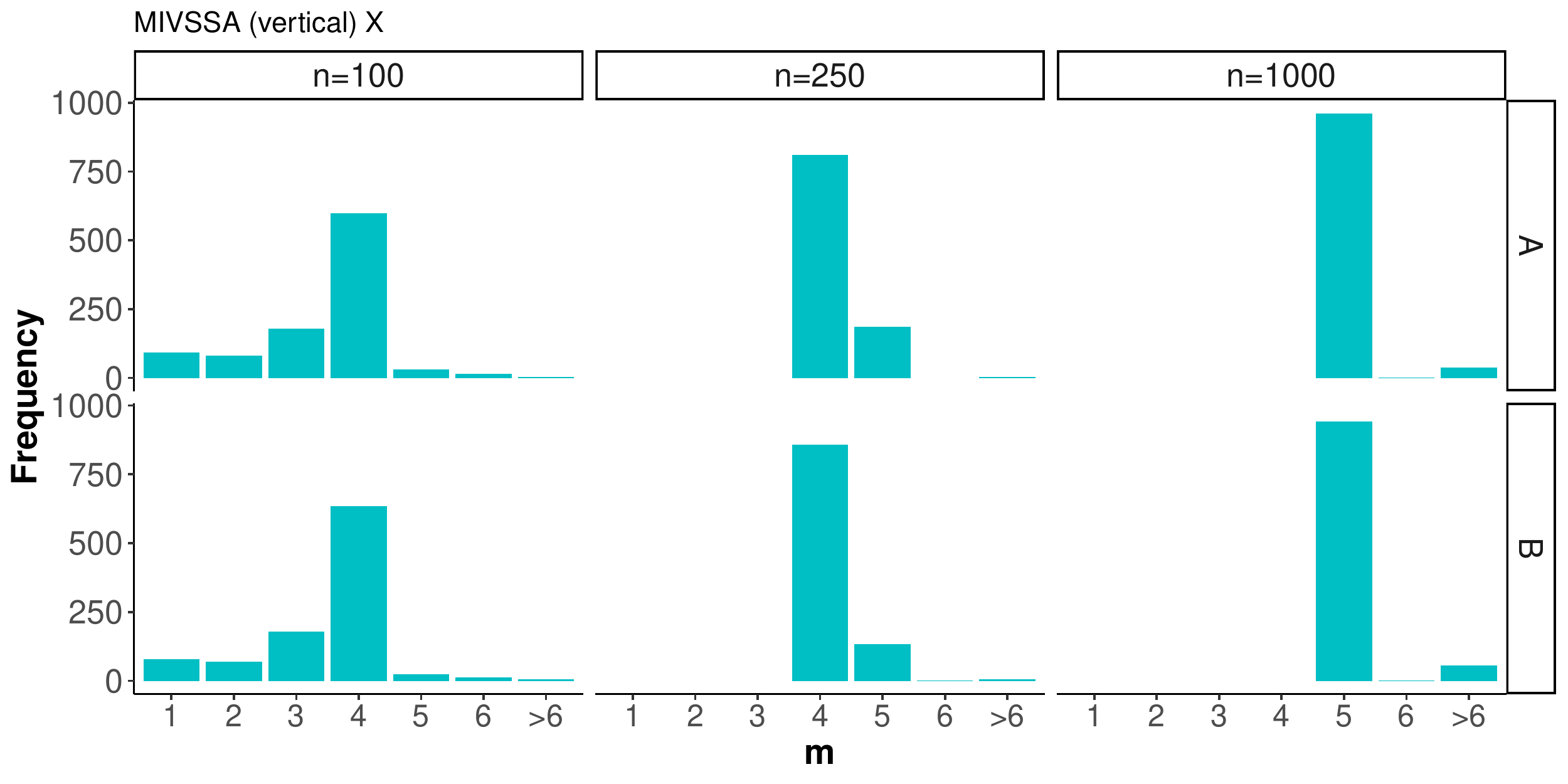}
\includegraphics[scale=0.33]{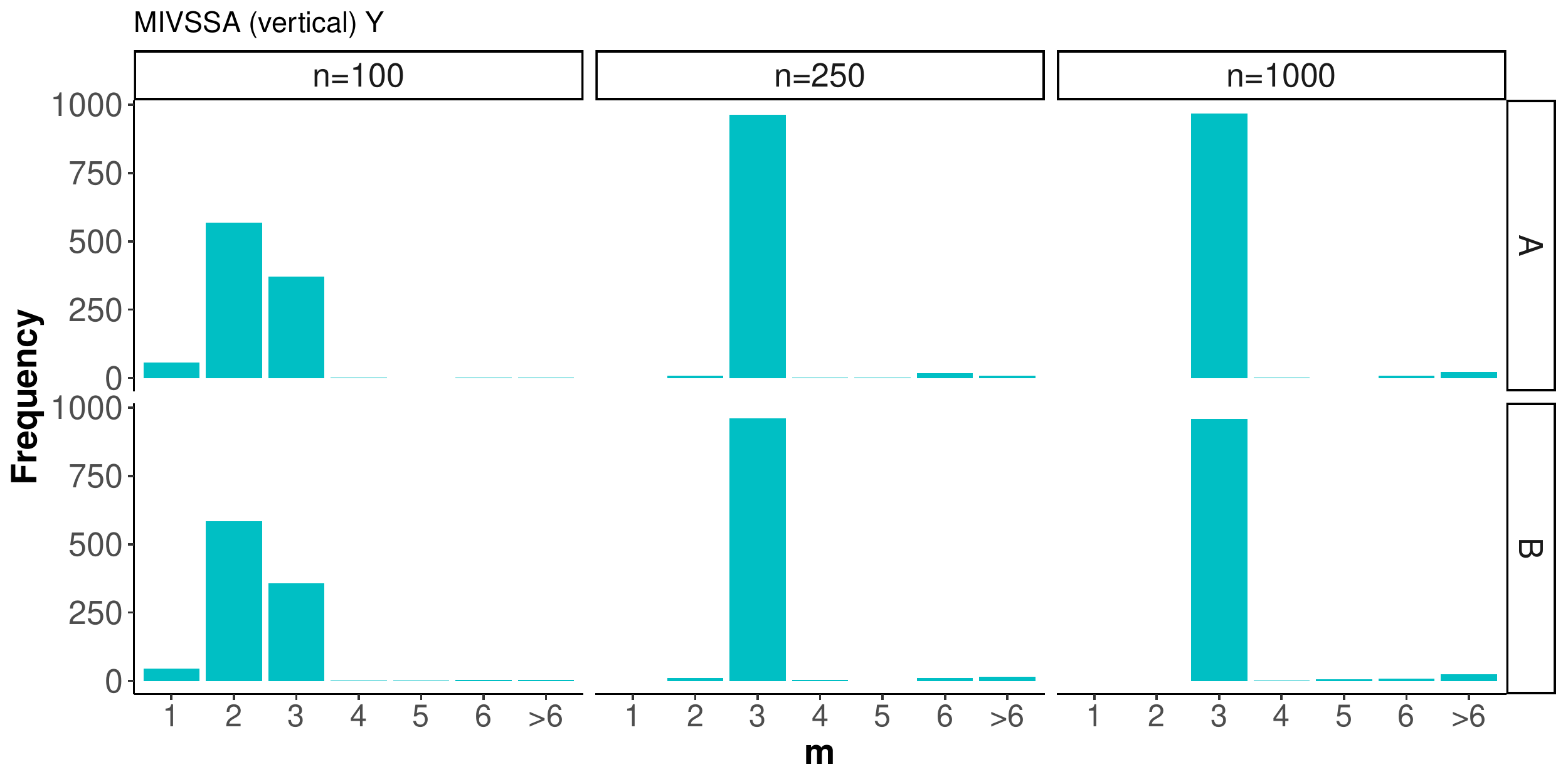}
\includegraphics[scale=0.33]{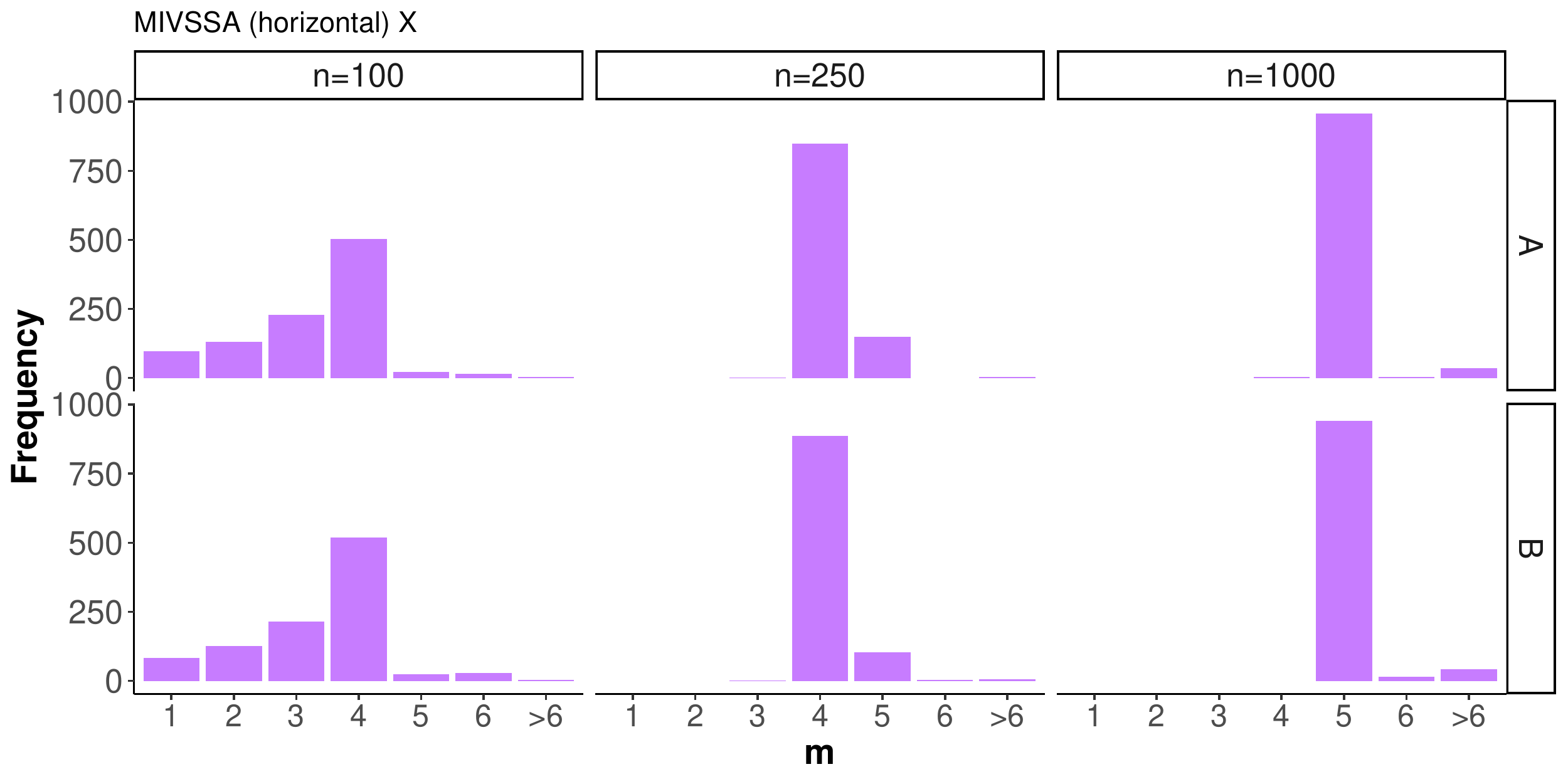}
\includegraphics[scale=0.33]{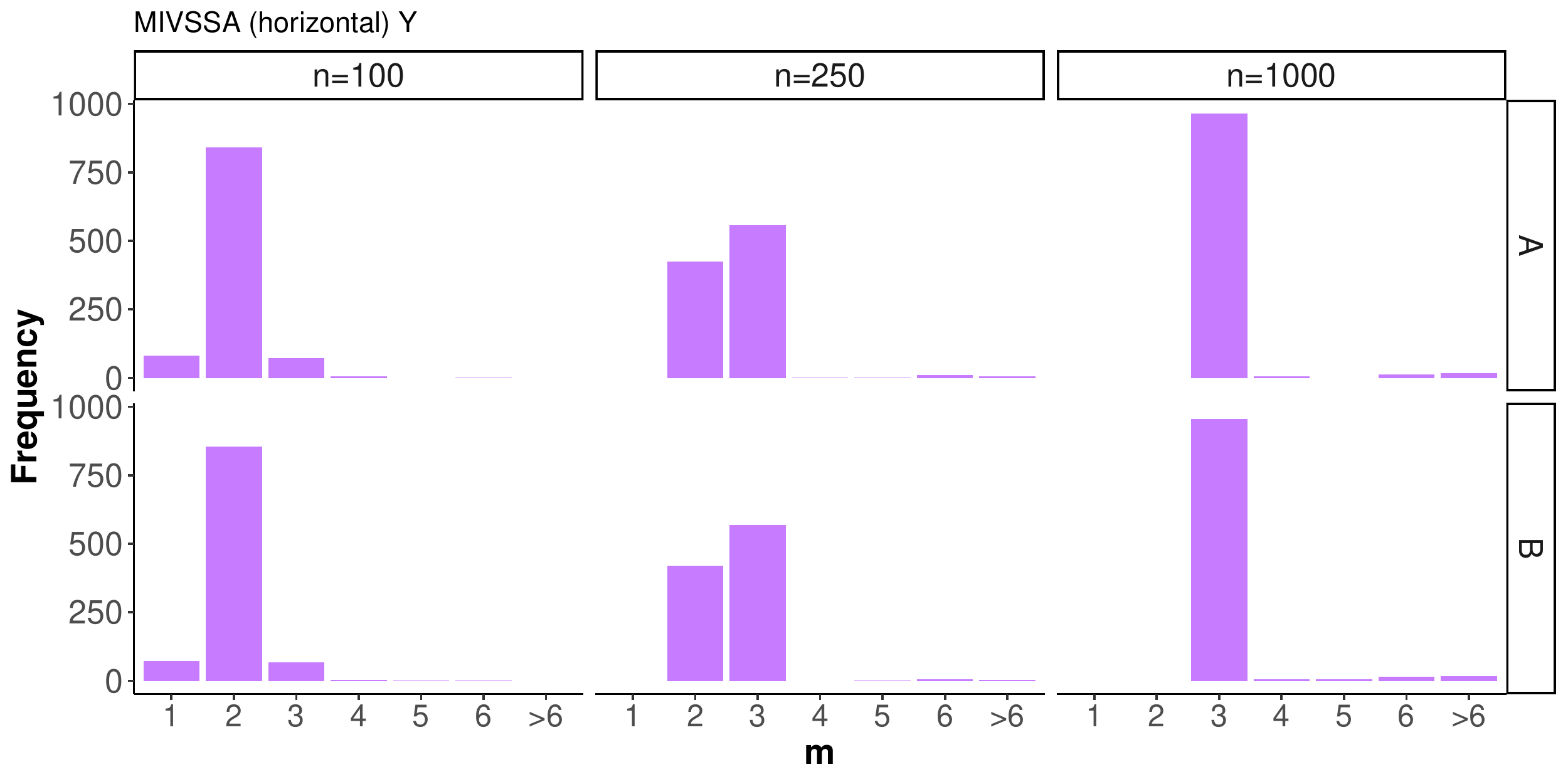}
\caption{\footnotesize \label{ercs} Number of ERC selected according to the periodogram-based of \cite{decarvalho2020} adapted for the context of interval-valued time series.}
\end{figure}

{As it can be seen from Figure~\ref{sce}, as the sample size increases the HE tends to decrease, regardless of the number of ERC. This thus indicates a better performance, from an Hausdorff residual perspective, of the proposed methods as the number of observations increases. We now switch gears and examine the periodogram-based criterion used for learning about the number of ERCs (see Appendix~A.2). Figure~\ref{ercs} displays the distribution of the number of ERCs selected with our automatic criterion on the Monte Carlo experiment. The joint analysis of Figures~\ref{sce} and \ref{ercs} suggests that our periodogram-based approach does a sensible job at learning about the number of ERCs as it tends to select a number of components (see Figure~\ref{ercs}) that closely follows the number of components achieving the lowest HE in the Monte Carlo simulation study (see Figure~\ref{sce}). 
}

\section{Interval Trendlines for Argentina Stock Market}\label{application}

\subsection{Data Description and Motivation for the Analysis}\label{motivation}
We now apply our methods so to learn about interval trendlines for the MERVAL index---the principal index of Argentina stock market. In Figure~\ref{fig:merval1} we depict the raw interval data series from Yahoo Finance corresponding to {weekly} minimum and maximum values of MERVAL ranging from January 1st 2016 to September 30th 2020. During the period of interest the economy of Argentina was impacted by several episodes of financial interest, and we will aim to examine how the trendlines of MERVAL reacted to those. Examples include: a) currency crisis started in (Q1 2018) \citep[][Section~4.1]{sturzenegger2019macri}, that involved the IMF (International Monetary Fund) intervention with 
a three--years lending program of USD 50bn \citep[][Section~4.2]{sturzenegger2019macri} approved in the end of (Q2 2018) \href{https://www.imf.org/en/News/Articles/2018/06/20/pr18245-argentina-imf-executive-board-approves-us50-billion-stand-by-arrangement}{(IMF Press release NO.18/245)} and later increased by USD 7bn 
\href{https://www.imf.org/en/News/Articles/2018/09/26/pr18362-argentina-imf-and-argentina-authorities-reach-staff-level-agreement}{(IMF Press release NO.18/362)} on (Q3 2018); b) the so-called PASO %8/ago/2019
 (primary elections in Argentina) whose surprising outcome (Q3 2019) has led to the imposition of foreign exchange controls %\citep{BBC}
 \href{https://www.bbc.com/news/business-49547189}{(BBC, press note)} and also a virtual sovereign debt default (Q4 2019) % Desde noviembre de 2019, después de que macri pierde las elecciones, se impula una ley de "reperfilamiento de deuda" (aprobada en diciembre de 2019). El link al decreto:
\href{https://www.argentina.gob.ar/normativa/nacional/decreto-49-2019-333534/texto}{(DNU 49/2019)} 
 ; c) a sovereign debt restructuring process between (Q1 2020) and (Q3 2020) \href{https://www.bloomberg.com/news/articles/2020-09-07/argentina-lifted-from-default-after-65-billion-restructuring}{(Bloomberg, press note)}; and d) COVID--19 lockdown over (Q1 2020--onwards) \href{https://www.bloomberg.com/news/articles/2020-03-20/argentina-orders-exceptional-lockdown-in-bid-to-contain-virus}{(Bloomberg, press note)}. The next section will employ the proposed methods and will assess how have the MERVAL trendlines reacted when those episodes took place.

\begin{figure}[h]
  \centering
\includegraphics[scale=0.66]{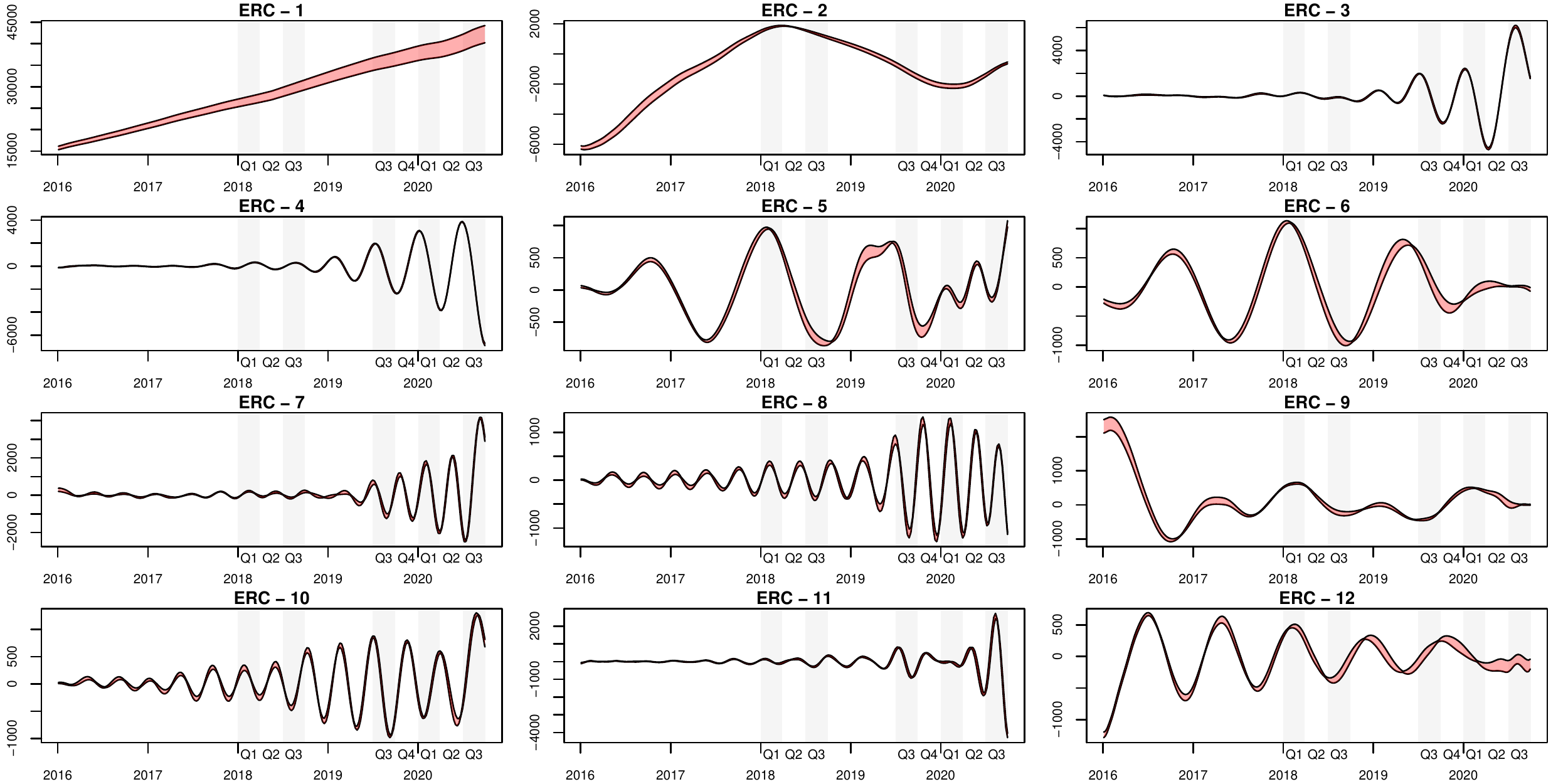}
\caption{\footnotesize First 12 ERCs of MERVA index obtained via IVSSA. The shaded areas represent episodes a--d from Section~\ref{motivation}\label{fig:comps}}
\end{figure}

\subsection{Modeling, Nowcasting, and Forecasting Interval Trendlines}
{Figure~\ref{fig:comps} depicts the first 12 ERCs of MERVAL obtained via the proposed decomposition methods. As it can be seen from the latter figure, the first components seem to correspond to movements associated with an interval drift, whereas the last few components seem to represent a cycle or noise; to draw a distinction between what ERCs that actually correspond to a drift, and which ones represent noise, we resort to an interval-valued version of the periodogram-based method of \cite{decarvalho2020}---which is discussed in Appendix~A.1, and whose performance has been examined in Section~\ref{simulation}.} In Figure~\ref{fig:merval1} we depict the MERVAL trendline corresponding to the IVSSA version of the proposed methods. To learn about the interval trendlines, we consider $l = \lceil(n+1)/2 \rceil = 125$ (as discussed on p.~\pageref{lrule}) and to learn about the number of components ($m$) we resort to our periodogram-based criterion. % To assess how the obtained interval trendlines change along with the number of components we also examine in this analysis other values of $m$, beyond $m^* = 31$ as obtained via our method. 
From a visualization viewpoint, perhaps a number components smaller than $m^* = 31$ is preferable, despite the overall good performance suggested by Section~\ref{simulation} of our criterion for selecting $m$. Keeping in mind this, and the fact that for forecasting the latter choice of $m$ may not be the most appropriate---as the interval-valued signal may follow the data too closely---we also use out-of-sample evaluations for selecting the values of $l$ and $m$. Roughly speaking, this is achieved by minimizing the {1Q (12 weeks) out-of-sample} Haussdorff residual {corresponding to models fitted over an expanding window}, and it yields $(l^{\star}, m^{\star}) = (80, 2)$; see Appendix~A.3 for details. Regardless of the value of $m$, as it can be seen from  Figure~\ref{fig:merval1}, the interval trendlines produced by our method have clear links with the episodes a--d mentioned in Section~\ref{motivation}. 
\begin{figure}[h]
  \centering
\includegraphics[scale=0.7]{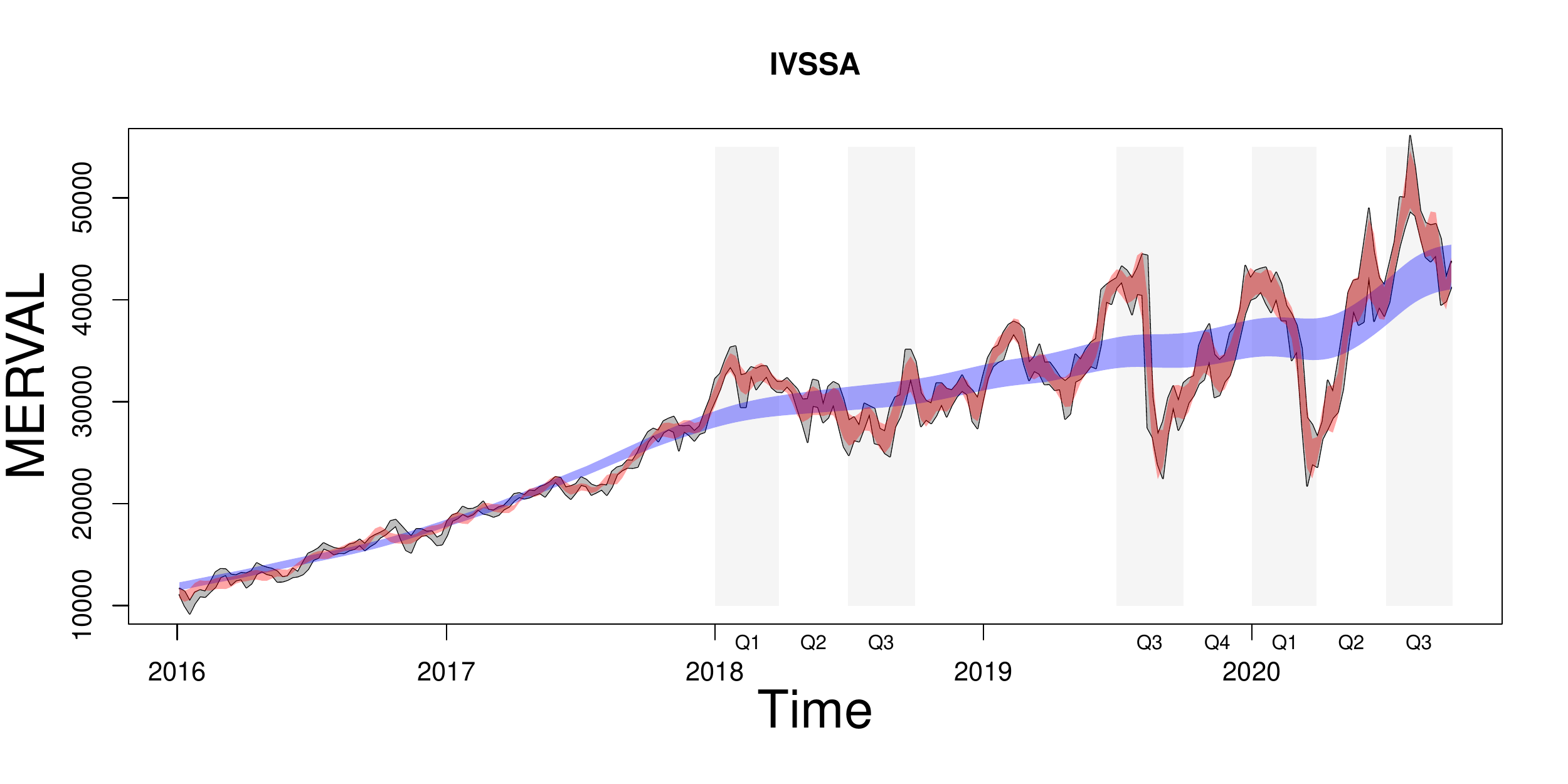}
\caption{\footnotesize `Post-mortem' interval trendlines: MERVAL interval-valued data defined by {weekly} minimum and maximum values of the index   ({\textcolor{gray}{$\medblacksquare$}}\textcolor{black}{)}, IVSSA interval trendline obtained with periodogram-based approach  ({\textcolor{red}{$\medblacksquare$}}\textcolor{black}{) and with Haussdorff residual-based approach ({\textcolor{dodgerblue}{$\medblacksquare$}}\textcolor{black}{). The shaded areas represent episodes a--d from Section~\ref{motivation}}\label{fig:merval1}}}
\end{figure}
%Yet, 

\begin{figure}
  \vspace{-2.5cm}
  \centering
\includegraphics[scale=0.6]{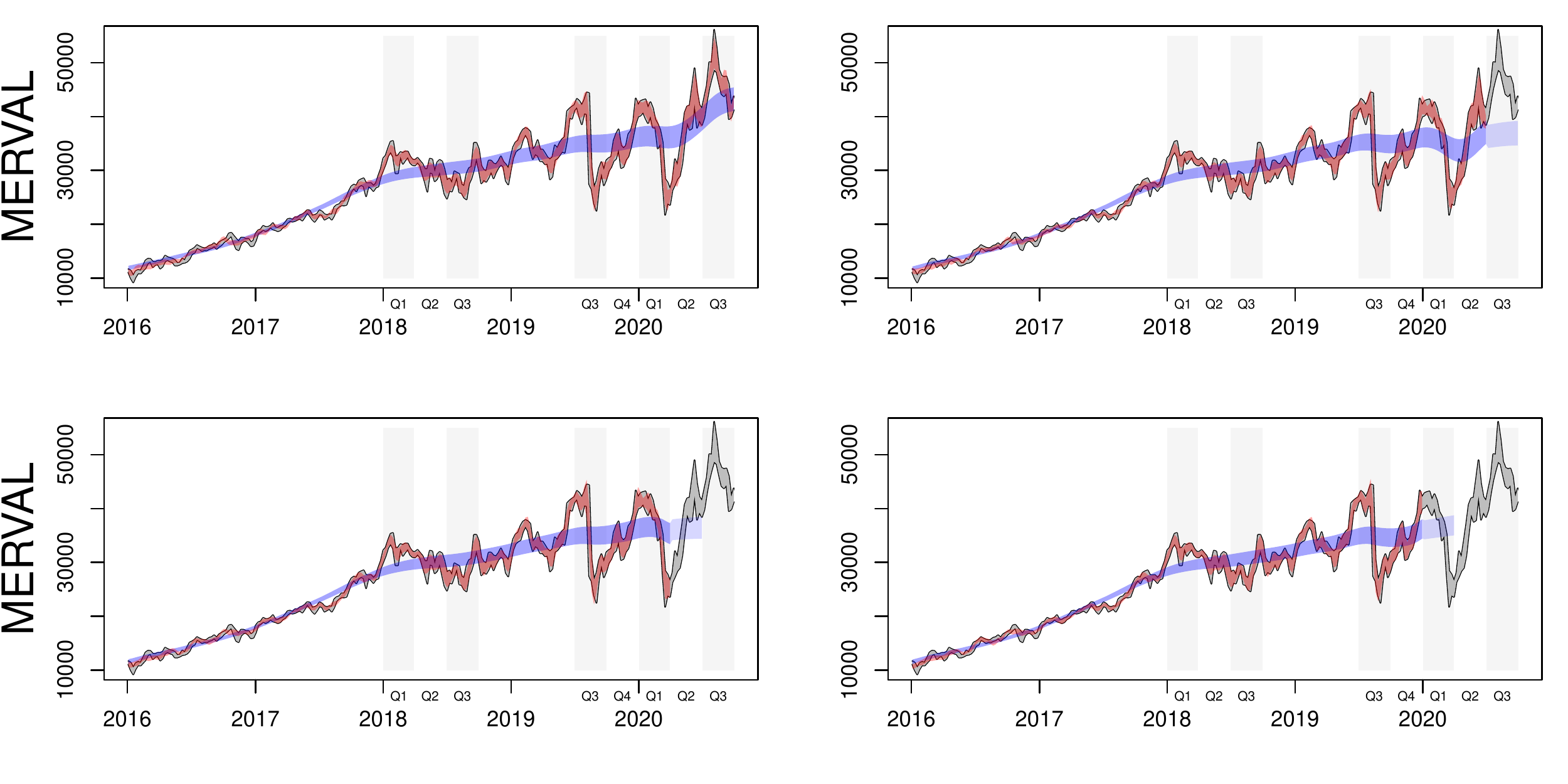}
\includegraphics[scale=0.6]{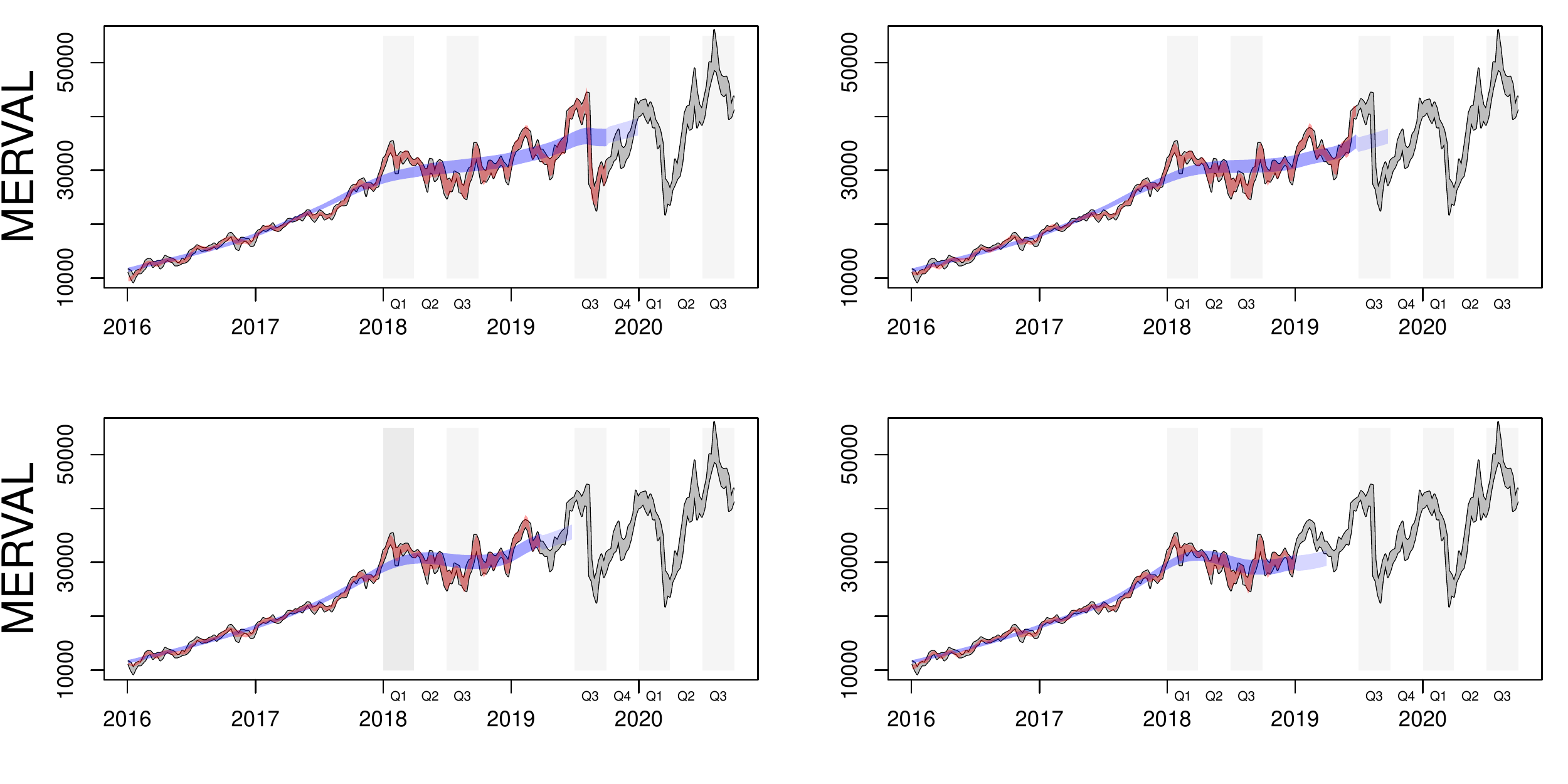}
\includegraphics[scale=0.6]{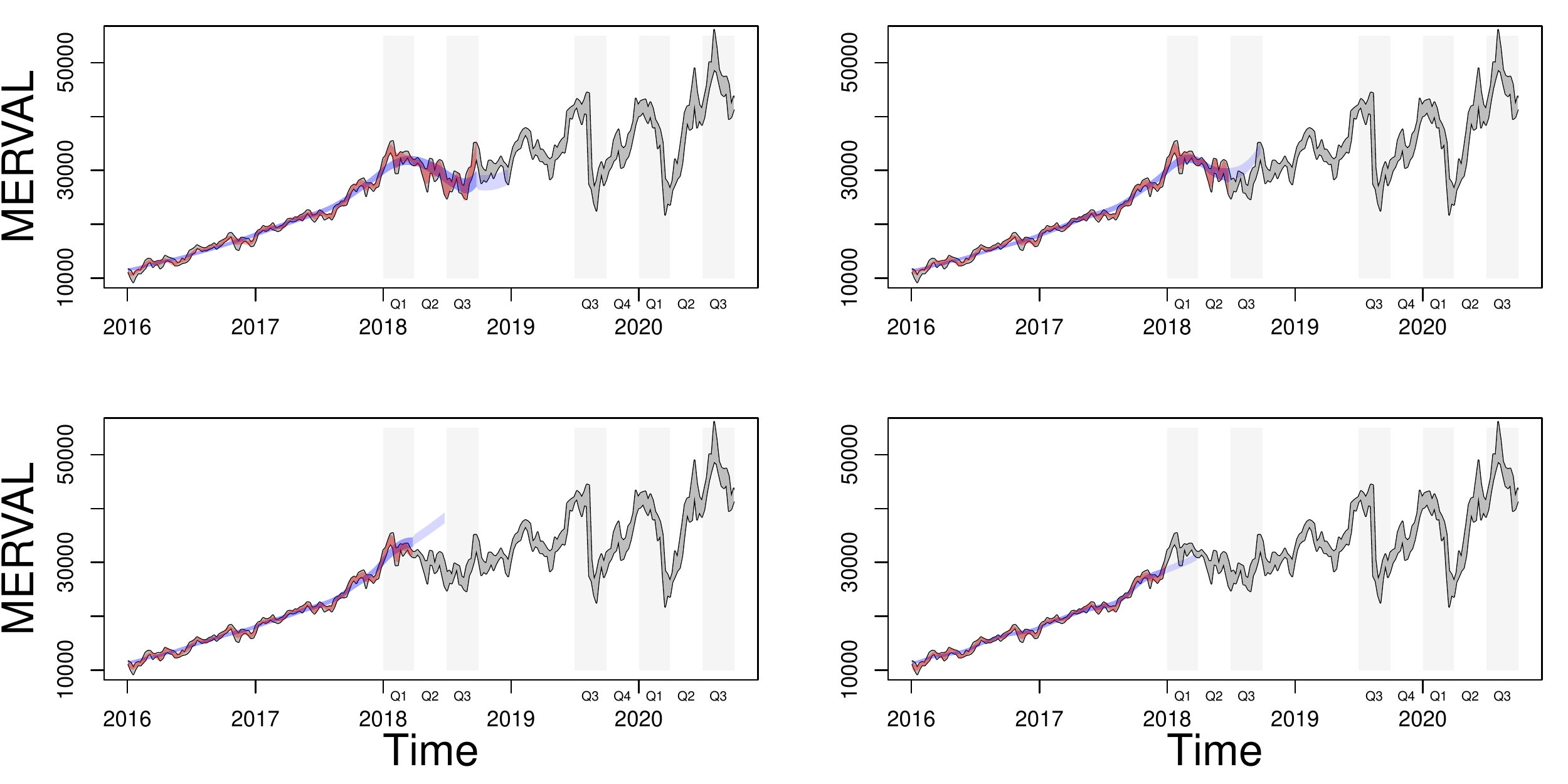}
\caption{\footnotesize Real-time analysis ranging from (Q3 2020) to (Q1 2018) along with MERVAL interval data ({\textcolor{gray}{$\medblacksquare$}}\textcolor{black}{)} and IVSSA interval trendline obtained with periodogram-based approach  ({\textcolor{red}{$\medblacksquare$}}\textcolor{black}{) and with Hausdorff residual-based approach } ({\textcolor{dodgerblue}{$\medblacksquare$}}\textcolor{black}{) along with the corresponding interval forecasts. The gray shaded areas represent episodes a--d from Section~\ref{motivation}.}\label{fig:merval2}}
\end{figure}

The interval trendlines depicted in Figure~\ref{fig:merval1} are a post-mortem in
the sense that they are based on the entire sample period. To assess how much the
trendlines produced by our method would be revised %---
when they are produced in real time, we conduct a real-time
analysis. In fields such as economics, the ability of a method to be
coherent over real-time---in the sense of not revising estimates once new data arrives---is key, and it has been a subject of wide
interest \citep[see for instance][and references therein]{orphanides2002}.
%
% The exercise is as follows, we use the sample period () as a training period, and we then compute real-time interval trendlines using our method. The result is depicted in Figure~\ref{fig:merval2}
We conduct a real-time analysis by sequentially removing the most recent quarters of data from the whole data set so to compute interval trendlines at the end of every quarter. The sequence of real-time interval trendlines is depicted in Figures~\ref{fig:merval2}. As it can be seen from the latter figure, our method does not revise substantially the produced interval trendlines; that is, the real-time interval trendlines (Figure~\ref{fig:merval2}) resemble those obtained from the post-mortem (Figure~\ref{fig:merval1}), thus suggesting a sensible real-time performance of our method. Such sturdy real-time performance is in line with what has been found for SSA for time series \citep{decarvalho2012e, decarvalho2020}---rather than for interval time series as examined here. The real-time analysis from Figure~\ref{fig:merval2}} also presents a sequence of out-of-sample forecasts that were obtained via the methods from Section~\ref{forecasting}. As it can be noticed from the latter figure, the out-of-sample forecasts obtained by the proposed method are reasonably in line with the true targets, thus suggesting a good forecast accuracy of the proposed methods in this real-time exercise.

\section{Closing Remarks}\label{discussion}

From a methodological outlook a main goal of this article was on
extending SSA-based methods to an interval time series context.  The
proposed extension is tailored for modeling a range of values over
time so to learn about interval-valued signals and to yield out-of-sample
forecasts of the said range of values.
To our knowledge this paper pioneers the development of statistical
decomposition methods for set-valued stochastic processes, and the proposed 
method can be used for decomposing an interval-valued time series into a string of components 
that can be interpreted as an interval-valued signal, cycle, or noise.
The proposed method coincides with standard SSA when the data of interest
are standard time series---rather than interval time series---and the 
multivariate extension of our method allows for combining
both time series and interval time series. 
Naively, one could think of applying standard multivariate SSA to
interval-valued data by treating each of the limits of an
interval as a vector as an alternative to the methodology proposed
herein; yet such naive multivariate SSA-based approach would not have
in mind the interval-valued nature of the data.

Some remarks on future research are in order.  It would seem natural
to use the spectral features learned via symbolic SSA so to cluster
or classify interval time series, or even to cluster according to these features;
given the recent applied relevance of clustering interval time series
\citep{maharaj2019}, we believe this could be a natural methodological
target for future investigation.  Another potential follow-up within the remit of this paper is the development
of decomposition methods for functional set-valued data, that would aim to extend
the methods proposed here to a continuous time setting. While
Functional Data Analysis \citep{ramsay2002,
  ferraty2006, ramsay2006, horvath2012}---i.e.~the analysis of data in the form of a continuous time stochastic
process---is a fast-evolving field, to our knowledge no
developments have been made on the statistical analysis of a set-valued version of functional 
data (i.e.~data in the form of a continuous time set-valued stochastic
process), nor have been devised decomposition methods for set-valued functional 
data. We leave such open problems for future analysis.

\section*{Appendix}\label{appendix}
\appendix \footnotesize 
\subsection*{A.1~Technical Details}\label{Proof}
\begin{proof}[Proof of Proposition~\ref{daver}]
  Our strategy is similar to that of \citet[][Proposition~6.3]{golyandina2001}. Since $\mathbf{H} = \{[\alpha_{i, j}, \beta_{i, j}]\}$ is an Hankel matrix of ordered pairs, it follows that it is constant across anti-diagonals, i.e, $[\alpha_{i, j}, \beta_{i, j}] = [g_{1,s}, g_{2,s}]$, for $i + j = s$ and some pair of numbers $(g_{1,s}, g_{2,s})$. Thus, it follows that 
  \begin{equation*}
    \begin{split}      
    \|\mathbf{Y} - \mathbf{H}\|_{\text{C}}^2  
    &= \sum_{i, j} \{(a_{i, j} - \alpha_{i, j})^2 + (b_{i, j} - \beta_{i, j})^2\} \\
    &= \sum_{i, j} (a_{i, j} - \alpha_{i, j})^2 + \sum_{i, j} (b_{i, j} - \beta_{i, j})^2 \\
    &= \sum_{s = 2}^{l + k} \sum_{i + j = s} (a_{i, j} - g_{1, s})^2 + 
    \sum_{s = 2}^{l + k} \sum_{i + j = s} (b_{i, j} - g_{2, s})^2, \\ 
    \end{split}
  \end{equation*}
    which is minimized for $g_{1,s} = {n_s}^{-1} \sum_{i + j = s} a_{i, j}$ and $g_{2,s} = {n_s}^{-1} \sum_{i + j = s} b_{i, j}$.
\end{proof}
\subsection*{A.2~Automatic Criterion to Choose the Number of ERC}
In this section we extend the automatic criterion to choose the number of ERC of \cite{decarvalho2020} to an interval-valued time series context. 
\subsubsection*{Groundwork on Spectral Analysis for Interval-Value Time Series}
Prior to introducing our criterion we need to lay the groundwork. The spectral density of an interval-valued time series  $\{y_t\equiv [a_t,b_t]\}_{t=1}^n$ is here defined as
\begin{equation}
  \label{eq:sdensity}
  f(\omega) = \frac{1}{2\pi} \sum_{h=-\infty}^{\infty} \gamma(h)e^{- ih\omega}, \quad \omega \in (-\pi,\pi], 
\end{equation}
where $\gamma(h) = \cov(y_t, y_{t - h})$ is the autocovariance function at lag $h \in \mathbb{N}$. 
The definition in \eqref{eq:sdensity} is motivated from the well-known relation between the autocovariance function and the spectral density \citep[][Proposition~10.1.2]{brockwell2002}. Further, we consider the interval residuals $\textbf{e} =\phi(\textbf{y} - \widetilde{\textbf{y}})=\{[e_1^L,e_1^U],\dots,[e_n^L,e_n^U]\}$
%=\{[e_i^L\equiv \min\{a_i-\widetilde{a}_i,b_i-\widetilde{b}_i\},e_i^U\equiv \max\{a_i-\widetilde{a}_i,b_i-\widetilde{b}_i\}]\}_{i=1}^n$ 
and the corresponding
spectral density plug-in estimator, which we will refer to as the periodogram for interval-valued time series, computed as follows: 
\begin{equation}
  \label{eq:periodogram}
  \widehat{f}(\omega_j) = \frac{1}{2\pi} \sum_{|h|\leq n-1}\widehat{\gamma}_e(h)e^{-ih\omega_j}, 
\end{equation}
where $\omega_{j}=2\pi j/n$ are the so-called Fourier frequencies, for
$j=1, \ldots, J = \lfloor (n-1)/2\rfloor$, with $\lfloor \cdot \rfloor$ denoting the floor function, and $\widehat{\gamma}_e(h)$ is the empirical autocovariance function of interval residuals which readily follows by adapting \citet[][Eq.~3]{billard2012}: 
\begin{equation*}
\widehat{\gamma}_{e}(h) = \frac{1}{6n}
\sum_{t=1}^{n-h}\big\{2e^L_te^L_{t+h} +e^L_te^U_{t+h} + e^U_te^L_{t+h} + 2e^U_te^U_{t+h}\big\}, \quad h=0,1,\dots,n-1.
\end{equation*}
Next, we show how \eqref{eq:periodogram} can be used for learning about the number of ERC.
\subsubsection*{Periodogram-Based Criterion for Learning about the Number of ERC}
Our periodogram-based criterion for learning about the number of ERC is tantamount to that of \cite{decarvalho2020}, but based on the periodogram for interval-valued time series in \eqref{eq:periodogram}. Below, $\omega_{j}=2\pi j / n$ are Fourier frequencies, for $j=1, \ldots, J = \lfloor n / 2\rfloor$, with $\lfloor \cdot \rfloor$ denoting the floor function. Formally, the method is as follows:} 
\begin{mdframed}[backgroundcolor=gray!45]
\begin{small}\label{targeted}
\noindent \textbf{Targeted grouping based on the Kolmogorov--Smirnov statistic} \\
Set $i = 1$ and execute the steps:
\begin{enumerate}[~~~Step 1.~]
\item Compute the interval residual vector $\textbf{e} =\phi(\textbf{y}-\widetilde{\textbf{y}})$, yield from the interval trendline based on $I = \{1,\dots, i\}$.
\item Compute the cumulative periodogram of $\textbf{e}$, and test the null hypothesis of white noise using the Kolmogorov--Smirnov test based on the statistic
  \begin{equation}
    \label{cumper}
    \sqrt{J} \max\{|C(\omega_j) - j/J|\}_{j = 1}^J, \quad C(\omega_j) = 
    \frac{\sum_{i = 1}^j \mathbb{I}(\omega_i)}{\sum_{i = 1}^J \mathbb{I}(\omega_i)}.
  \end{equation}
\end{enumerate}
If the null is rejected %and $\int_0^{\pi} C(\omega) \, \dif \omega > \pi / 2$, 
then increment $i$ and repeat Steps 1 and 2. Otherwise stop.
\end{small}
\end{mdframed}\footnotesize 
In words, the method sequentially adds components until there is evidence from the cumulative periodogram of the interval residuals suggesting that the interval residuals constitutes white noise.

\subsection*{A.3~Hausdorff Residual-Based  Criterion for Learning about Window Length and Number of ERC for Forecasting}

Let $\widehat{y}_t \equiv \widehat{y}_t(\mathscr{D}_w,l,m)$ be the forecast obtained with IVSSA using the method from Section~\ref{forecasting} with the training data set $\mathscr{D}_{w} = (y_1,\dots,y_{w})$, for $1 < w \leq n$, and with parameters $(l,m)$. To learn about the smoothing parameters $(l, m)$ in the context of forecasting $p$-steps ahead of period $w_0$, we solve the minimization problem 
$$(l^{\star},m^{\star}) = \arg \underset{(l,m)}{\min} \sum_{w=w_0}^{n-p} \sum_{t=w+1}^{w+p} D_{\mathrm {H} }(y_t,\widehat{y}_t),$$
where $D_{\mathrm {H} }$ is the Hausdorff distance. (For example, for the forecasts depicted in Figure~\ref{fig:merval2}, we consider $w_0=104$---the weeks corresponding to years 2016 and 2017 in MERVAL data---and $p=12$, i.e.~we choose $(l,m)$ so to maximize 1Q forecasting accuracy.)
\section*{Acknowledgments} \label{acknowledgements}\footnotesize
The research was partially funded by the project INTERSTATA (\textbf{Inter}disciplinary \textbf{Stat}istics in \textbf{A}ction) from the International Research and Partnership Fund (Developing Countries), and from FCT (Funda\c c\~ao para a Ci\^encia e a Tecnologia, Portugal), through the projects PTDC/MAT-STA/28649/2017 and UID/MAT/00006/2019.

\bibliographystyle{LaTeX/BibtexStyles/asa2.bst} 
\bibliography{LaTeX/library.bib}

\begin{thebibliography}{27}
\newcommand{\enquote}[1]{``#1''}
\expandafter\ifx\csname natexlab\endcsname\relax\def\natexlab#1{#1}\fi

\bibitem[{Billard(2007)}]{billard2007}
Billard, L. (2007), \enquote{Dependencies and variation components of symbolic
  interval-valued data,} in \textit{Selected Contributions in Data Analysis and
  Classification}, Springer, pp. 3--12.

\bibitem[{Billard and Le-Rademacher(2012)}]{billard2012}
Billard, L. and Le-Rademacher, J. (2012), \enquote{Principal component analysis
  for interval data,} \textit{Wiley Interdisciplinary Reviews: Computational
  Statistics}, 4, 535--540.

\bibitem[{Brockwell and Davis(2002)}]{brockwell2002}
Brockwell, P.~J. and Davis, R.~A. (2002), \textit{Time {{Series}}: {{Theory}}
  and {{Methods}}}, New York: {Springer}.

\bibitem[{{de Carvalho} and Martos(2018)}]{dm2018}
{de Carvalho}, M. and Martos, G. (2018), \textit{ASSA: Applied Singular
  Spectrum Analysis}, {R} package version 1.0.

\bibitem[{de~Carvalho and Martos(2020)}]{decarvalho2020}
de~Carvalho, M. and Martos, G. (2020), \enquote{Brexit: Tracking and
  disentangling the sentiment towards leaving the EU,} \textit{International
  Journal of Forecasting}, 36, 1128--1137.

\bibitem[{{de Carvalho} et~al.(2012){de Carvalho}, Rodrigues, and
  Rua}]{decarvalho2012e}
{de Carvalho}, M., Rodrigues, P.~C., and Rua, A. (2012), \enquote{Tracking the
  {{US}} business cycle with a singular spectrum analysis,} \textit{Economics
  Letters}, 114, 32--35.

\bibitem[{Ferraty and Vieu(2006)}]{ferraty2006}
Ferraty, F. and Vieu, P. (2006), \textit{Nonparametric Functional Data
  Analysis: Theory and Practice}, New York: {Springer}.

\bibitem[{Golyandina et~al.(2001)Golyandina, Nekrutkin, and
  Zhigljavsky}]{golyandina2001}
Golyandina, N., Nekrutkin, V., and Zhigljavsky, A.~A. (2001), \textit{Analysis
  of Time Series Structure: SSA and Related Techniques}, Boca Raton, FL:
  Chapman and Hall/CRC.

\bibitem[{Golyandina and Zhigljavsky(2013)}]{golyandina2013}
Golyandina, N. and Zhigljavsky, A. (2013), \textit{Singular Spectrum Analysis
  for Time Series}, New York: Springer.

\bibitem[{Gonz{\'a}lez-Rivera and Arroyo(2012)}]{gonzalez2012}
Gonz{\'a}lez-Rivera, G. and Arroyo, J. (2012), \enquote{Time series modeling of
  histogram-valued data: The daily histogram time series of S\&P500 intradaily
  returns,} \textit{International Journal of Forecasting}, 28, 20--33.

\bibitem[{Gonz{\'a}lez-Rivera and Lin(2013)}]{gonzalez2013}
Gonz{\'a}lez-Rivera, G. and Lin, W. (2013), \enquote{Constrained regression for
  interval-valued data,} \textit{Journal of Business \& Economic Statistics},
  31, 473--490.

\bibitem[{Hassani and Mahmoudvand(2013)}]{hassani2013}
Hassani, H. and Mahmoudvand, R. (2013), \enquote{Multivariate singular spectrum
  analysis: {{A}} general view and new vector forecasting approach,}
  \textit{International Journal of Energy and Statistics}, 1, 55--83.

\bibitem[{Horv{\'a}th and Kokoszka(2012)}]{horvath2012}
Horv{\'a}th, L. and Kokoszka, P. (2012), \textit{Inference for {{Functional
  Data}} with {{Applications}}}, vol. 200, New York: {Springer}.

\bibitem[{Khan and Poskitt(2017)}]{khan2017}
Khan, M. A.~R. and Poskitt, D. (2017), \enquote{Forecasting stochastic
  processes using singular spectrum analysis: Aspects of the theory and
  application,} \textit{International Journal of Forecasting}, 33, 199--213.

\bibitem[{Kisielewicz(2013)}]{kisielewicz2013}
Kisielewicz, M. (2013), \textit{Set-Valued Stochastic Processes}, New York, NY:
  Springer New York, pp. 67--102.

\bibitem[{Le-Rademacher and Billard(2012)}]{le2012}
Le-Rademacher, J. and Billard, L. (2012), \enquote{Symbolic covariance
  principal component analysis and visualization for interval-valued data,}
  \textit{Journal of Computational and Graphical Statistics}, 21, 413--432.

\bibitem[{Lin and Gonz{\'a}lez-Rivera(2016)}]{lin2016}
Lin, W. and Gonz{\'a}lez-Rivera, G. (2016), \enquote{Interval-valued time
  series models: Estimation based on order statistics exploring the agriculture
  marketing service data,} \textit{Computational Statistics \& Data Analysis},
  100, 694--711.

\bibitem[{Maharaj et~al.(2019)Maharaj, Teles, and Brito}]{maharaj2019}
Maharaj, E.~A., Teles, P., and Brito, P. (2019), \enquote{Clustering of
  interval time series,} \textit{Statistics and Computing}, 29, 1011--1034.

\bibitem[{Mahmoudvand and Rodrigues(2018)}]{mahmoudvand2018}
Mahmoudvand, R. and Rodrigues, P.~C. (2018), \enquote{A new parsimonious
  recurrent forecasting model in singular spectrum analysis,} \textit{Journal
  of Forecasting}, 37, 191--200.

\bibitem[{Orphanides and Van~Norden(2002)}]{orphanides2002}
Orphanides, A. and Van~Norden, S. (2002), \enquote{The Unreliability of
  Output-Gap Estimates in Real Time,} \textit{Review of Economics and
  Statistics}, 84, 569--583.

\bibitem[{{R Development Core Team}(2016)}]{rdevelopmentcoreteam2016}
{R Development Core Team} (2016), \textit{R: {{A Language}} and {{Environment}}
  for {{Statistical Computing}}}, Vienna, Austria: {R Foundation for
  Statistical Computing}.

\bibitem[{Ramsay(2006)}]{ramsay2006}
Ramsay, J.~O. (2006), \textit{Functional {{Data Analysis}}}, New York: {Wiley}.

\bibitem[{Ramsay and Silverman(2002)}]{ramsay2002}
Ramsay, J.~O. and Silverman, B.~W. (2002), \textit{Applied {{Functional Data
  Analysis}}: {{Methods}} and {{Case Studies}}}, vol.~77, {Citeseer}.

\bibitem[{Rodrigues and {de Carvalho}(2013)}]{rodrigues2013}
Rodrigues, P.~C. and {de Carvalho}, M. (2013), \enquote{Spectral modeling of
  time series with missing data,} \textit{Applied Mathematical Modelling}, 37,
  4676--4684.

\bibitem[{Rodrigues and Salish(2015)}]{rodrigues2015}
Rodrigues, P.~M. and Salish, N. (2015), \enquote{Modeling and forecasting
  interval time series with threshold models,} \textit{Advances in Data
  Analysis and Classification}, 9, 41--57.

\bibitem[{Sturzenegger(2019)}]{sturzenegger2019macri}
Sturzenegger, F. (2019), \enquote{Macri's Macro: The elusive road to stability
  and growth,} \textit{Brookings Papers on Economic Activity}, 2019, 339--436.

\bibitem[{Wang et~al.(2016)Wang, Zhang, and Li}]{wang2016}
Wang, X., Zhang, Z., and Li, S. (2016), \enquote{Set-valued and interval-valued
  stationary time series,} \textit{Journal of Multivariate Analysis}, 145,
  208--223.

\end{thebibliography}

\end{document}